\documentclass[twocolumn]{aastex631}

\received{October 18, 2023}
\revised{April 19, 2024}
\accepted{May 5, 2024}
\published{July 18, 2024}

\begin{document}

\title[Star-forming Es in MaNGA]{Structure and Kinematics of Star-forming Elliptical Galaxies in SDSS-MaNGA}

\author[0009-0002-2173-0953]{Pralay Biswas}
\affiliation{National Centre for Radio Astrophysics, Tata Institute of Fundamental Research, Post Bag 3, Ganeshkhind, Pune 411007, India}
\email{pbiswas@ncra.tifr.res.in}
\author[0000-0002-1345-7371]{Yogesh Wadadekar}
\affiliation{National Centre for Radio Astrophysics, Tata Institute of Fundamental Research, Post Bag 3, Ganeshkhind, Pune 411007, India}
\email{yogesh@ncra.tifr.res.in}






\begin{abstract}

Using ‘spatially’ resolved spectroscopy, we investigated the characteristics and different modes of formation of stars in elliptical galaxies. We identified an unusual population of 59 star-forming elliptical (SF-E) galaxies in SDSS-MaNGA, our primary sample. To identify these rare star-forming ellipticals, we combined GSWLC-A2 containing outputs of stellar population synthesis models with morphological results from the deep-learning catalog and resolved and integrated properties from the MaNGA Pipe3D value-added catalog. We have also constructed two control samples of star-forming spirals (SF-Sps; 2419 galaxies) and quenched ellipticals (Q-Es; 684 galaxies) to compare with our primary sample of SF-Es. H$\alpha$ emission line flux of SF-Es is similar to spiral galaxies. The D4000 spectral index indicates that SF-Es have a mixture of old and young stellar populations. Mass-weighted stellar age and metallicity for the SF-Es are lower than the Q-Es and 67\% of stellar- and gas- velocity maps of the primary sample show signs of kinematic disturbance. All of these indicate that SF-Es have acquired metal-poor gas through recent mergers or interactions with other galaxies and are forming a new generation of stars. Further, we subdivide our primary sample of SF-Es into four classes based on their bulge to total luminosity ratio $(B/T)$ and spin parameter $\lambda_{re}$. These four classes have their distinct evolutionary history and modes of formation. Based on these results, we suggest that the Hubble diagram does not accurately capture galaxy evolution processes, and we need a revised morphology diagram like the comb morphology diagram to get a complete picture of the galaxy evolution processes.

\end{abstract}



\section{Introduction} \label{sec: introduction}

Galaxies are gravitationally bound systems of stars, interstellar gas, dust, and dark matter. Galaxies are characterized by several global properties: luminosity, stellar mass (M$_{\ast})$, morphology, size, star-formation rate (SFR), molecular/atomic/ionized gas content, environmental density, bulge to total luminosity ratio $(B/T)$, etc. These properties, when measured quantitatively, constitute a multidimensional parameter space. The distribution of galaxies in this space is not random, as many of these parameters are correlated with each other and manifest as scaling relations. Some of the well-known scaling relations are the color-magnitude relation \citep{Sandage78}, the Tully-Fisher relation \citep{Tully77}, the Fundamental Plane \citep{Djorgovski87, Dressler87} and the size-magnitude relation \citep{Romanishin86}. By studying these correlations while paying particular attention to galaxies that are outliers from the overall trend, we can gain insight into the galaxy formation and evolution processes responsible for these correlations. 

With large-area galaxy surveys, like the Sloan Digital Sky Survey \citep[SDSS;][]{York00}, researchers found that, in the nearby Universe, there is a bimodal distribution of galaxies on the color-magnitude diagram \citep{Strateva01, Kauffmann03,
Baldry04, Baldry06}. There are the ‘blue cloud’ galaxies with active star formation and the ‘red sequence’ galaxies, which evolve passively, with little or no current star formation. The galaxies in the region between these two populations are termed the ‘green valley’ galaxies \citep{Wyder07}. Green valley galaxies are believed to be undergoing quenching of star formation \citep{Salim07} and are in the transition zone between the blue cloud and the red sequence \citep{Wyder07, Schiminovich07, Mendez11}. Based on visual morphologies, galaxies are broadly classified as spirals (Sps), lenticulars (S0s), ellipticals (Es), and irregulars \citep{Hubble26}. Many physical parameters discussed above correlate strongly with galaxy morphology, e.g. the SFR, which is high for Sps and low for Es \citep{Kennicutt98}.

Further, the star-forming main sequence (SFMS) refers to the tight relationship observed between the M$_*$ and SFR of star-forming galaxies. This SFMS is observed in the local \citep{Brinchmann04, Salim07} as well as the high-redshift Universe up to $z\sim5$ \citep{Daddi07, Elbaz07, Santini17}. Galaxies can deviate from the SFMS due to various physical processes, leading to the quenching of star formation. The exact physical processes behind the quenching of galaxies are unknown and often debated in the literature. Researchers invoke various mechanisms to explain quenching in star-forming galaxies, and morphological quenching is one of the strong candidate mechanisms \citep{Martig09}. Late-type galaxies (LTGs: Sps) belong to the blue sequence and are highly star forming. On the contrary, early-type galaxies (ETGs: Es and S0s) are red and quenched. ETGs are massive, reside in dense environments, have low gas content, often harbor strong AGNs (active galactic nuclei) and have a merger-rich evolutionary history. All of these characteristics could contribute to star formation quenching in these galaxies.

Early-type galaxies are usually passive and red. However, a number of studies have recently uncovered an actively star-forming population of early-type galaxies. \cite{Schawinski09} reported 204 blue early-type galaxies having SFR between 0.5 and 50 M$_{\odot} yr^{-1}$ in SDSS Data Release 6 (DR6) at $0.02<z<0.05$. For visual morphology, they used the Galaxy Zoo catalog and $u-r$ colors as a proxy for the SFR. They also found that blue early-type galaxies reside in lower density regions than early-type red galaxies, constituting nearly 5\% of the early-type population in the nearby universe. Although this particular class of galaxy is scarce, their presence raises pointed questions about our understanding of early-type galaxies.

There have been a number of efforts in the literature to try to understand the origin and characteristics of star-forming ETGs. \citet{George15}, and \citet{George17} studied a sample of 55 star-forming early-type galaxies and argued that recent minor mergers or interactions with gas-rich nearby galaxies could be responsible for the star formation in these galaxies. Once the acquired gas is depleted, these galaxies will evolve as standard passive ellipticals. In another study using $NUV-r$ color $(NUV-r \leq 5.0$) as the selection criterion for star-forming galaxies, \citet{Jeong21} identified recent star formation in the early-type galaxies and found that 7.7\% of the sample galaxies have recent star formation. Investigation of the metal abundance in these galaxies shows that early-type galaxies with recent star formation are younger and more metal poor than quiescent early-type galaxies. This suggests that these galaxies acquired metal-poor gas due to recent mergers or interactions with other galaxies.

\citet{Kaviraj09} compared synthetic photometry from numerical simulations of minor mergers with UV colors from nearby $(0.05 < z < 0.06)$ early-type galaxies. The simulations were performed with merger progenitors with reasonable assumptions about ages, dust properties, and metallicities. The study indicated that minor mergers are a potent mechanism behind the large UV scatter and low-level star formation in early-type galaxies.
 
There are also a few studies on star-forming S0s trying to understand the different formation mechanisms of these galaxies. In the SDSS-MaNGA \citep[Sloan Digital Sky Survey-Mapping Nearby Galaxies at Apache Point Observatory;][]{Bundy15, Blanton17}, \citet{Rathore22} identified a sample of 120 star-forming S0s (SF-S0s). They found that the population of SF-S0s is abundant at lower stellar masses but sharply decreases above $10^{10}$ $\rm{M}_{\odot}$. SFR surface density and specific star formation rate (sSFR) radial profiles show that the star formation in SF-S0s is comparable to the star-forming spirals (SF-Sps) but centrally dominated and decreases with the distance from the center. By visually inspecting the stellar- and gas- velocity maps, they found that more than 50\% SF-S0s are kinematically unsettled, possibly indicating signs of disturbance due to recent mergers rejuvenating the star formation. Another study by \citet{Xu21} tried to understand the environmental dependence of the formation mechanism of SF-S0s. They argued that the SF-S0s in the groups are possibly the successors of fading spirals. However, in the field, S0s could have ignited their star formation through minor mergers providing metal-poor gas.

From the above discussion, it is clear that any sample of star-forming ETGs may not be a homogeneous class of objects. Different formation mechanisms could be responsible, and their relative abundance and global properties depend strongly on several parameters like stellar mass and environment. In this work, we study a rare class of starforming ellipticals (SF-Es) in the nearby universe $(z < 0.1)$ using the SDSS-MaNGA Integral Field Spectroscopy (IFS) survey. Being an IFS survey, MaNGA provides spatially resolved spectroscopy across the 2D image of the galaxies. This enables us to study different properties of the galaxies as well as their stellar population on spatially resolved scales.  For a deeper understanding, we also compare the similarities and differences between star-forming ellipticals and two control samples of star-forming spirals and typical quenched ellipticals. This comparison could shed light on the different modes of formation of star-forming ellipticals. 

This paper is organized as follows: We describe the sample selection in Section \ref{sec: sample selection And Characterization} along with the MaNGA survey, the different catalogs used, the selection criteria, and some of the global properties of our three samples. In Section \ref{sec: results}, we present our main results. The results of the whole work and the different formation scenarios of the SF-Es are discussed in Section \ref{sec: discussion}. We summarize the study and list the main conclusions in Section \ref{sec: summary}. Throughout, we use the $\Lambda$CDM cosmology with parameters H$_0=67.8 \: \: {\rm km} \: {\rm s}^{-1} \: {\rm Mpc}^{-1}$, $\Omega_\Lambda = 0.69$ and $\Omega_m = 0.31$ \citep[Planck13 cosmology;][]{Planck13}.

\section{Sample Selection and Characterization} \label{sec: sample selection And Characterization}
This section describes the sample selection criteria for the primary sample and the two control samples. We describe briefly the MaNGA survey, the Pipe3D pipeline, and all the catalogs we use in this study. We also briefly discuss each sample and some of its global properties.

\subsection{The data and sample selection criteria} \label{sec:sample_overview}

\subsubsection{The SDSS-MaNGA galaxy survey} \label{sec:SDSS-MANGA}
The MaNGA survey is an SDSS-IV project \citep{Blanton17} that has observed over 10,000 galaxies from 2014 to 2020 \citep{Bundy15, Drory15, Abdurro'uf22}. It uses the SDSS 2.5 m telescope at APO \citep[Apache Point Observatory;][]{Gunn06} and the BOSS \citep[Baryon Oscillation Sky Survey;][]{Smee13} spectrographs with continuous wavelength coverage from 3600 to 10300 \text{\AA}. This is an IFS survey with fibers bundled into a distribution of several Integral Field Units (IFUs) of different sizes, which are tightly packed hexagonal arrays of fibers that deliver spatially resolved spectroscopy of each galaxy. The MaNGA survey has complicated selection criteria driven by scientific reasons to enhance the utility of the sample galaxies. The total MaNGA sample consists of three main subsamples. The first one is the Primary sample, which has a flat distribution in K-corrected $i$-band absolute magnitude ($M_i$) with coverage to 1.5 $R_e$ ($r$-band effective radius along the major axis). The Secondary sample also has a flat distribution in $M_i$ but is covered by the IFUs to 2.5 $R_e$. Finally, a sample of galaxies were selected to add galaxies in the underrepresented regions of $NUV - i$ versus $M_i$ color-magnitude plane in the primary sample called the Color-Enhanced sample with coverage to 1.5 $R_e$. The Primary and the Color-Enhanced sample together form the Primary+ sample, which is two-thirds of the whole sample, and the remaining one-third is the Secondary sample. Since the survey is designed to have a flat number density in $M_i$ (proxy of the stellar mass) and flattened in color for the Color-Enhanced sample, it has strong selection biases. It also has technical biases because of the different IFU sizes and constraints on their allocations. These biases affect the statistical analysis of any population of galaxies. To carry out statistical analysis, it becomes essential to use the volume weight provided for each galaxy \citep{Wake17}. These weights can be used for correcting back to a volume-limited sample to remove these biases. In this paper, whenever computing any statistical quantity like mean or median for a sample of galaxies, we have used these volume correction factors and referred to the quantity as weighted mean or weighted median. All the histograms shown are also weighted using these weights.

\subsubsection{Pipe3D pipeline} \label{sec:Pipe3D}
All MaNGA galaxies were reduced by Pipe3D \citep{SanchezI16}, a fitting pipeline to analyze IFS data  \citep{Sanchez18}. Apart from MaNGA, Pipe3D was also used to process the data from other large IFS surveys such as the Sydney-AAO Multi-object Integral-field spectrograph (SAMI) and Calar Alto Legacy Integral Field Area (CALIFA) surveys \citep{CALIFA12, SanchezI16, SanchezII16, Allen15, Croom21}. Also, multiple previous works with the MaNGA survey have used Pipe3D outputs. In this subsection, we will briefly discuss Pipe3D and its outputs; a complete description of the software, MaNGA data reduction, and outputs can be found in \citet{SanchezI16, SanchezII16, Sanchez18}. To increase the signal-to-noise ratio (S/N) of the stellar continuum, a binning of the spatial pixels (spaxels) is performed, called the Voronoi tessellation \citep{Cappellari03}. This technique aims to get a uniform S/N per bin by performing adaptive spatial binning of the IFS observations. This is an important requirement for the proper analysis of IFS data. At the beginning, Pipe3D analyzes the stellar population and the strong emission lines, then separately analyzes the weak emission lines. Each bin is fitted with a template of simple stellar populations (SSPs) for the stellar population analysis, given the dust attenuation and kinematics. The fitted continuum is subtracted from the full spectrum to get a data cube with only emission lines. Gaussian profiles are used to fit the strong emission lines, and these model emission line profiles are subtracted from the original spectrum to remove the effect of the strong emission lines. A stellar library is used to fit these strong emission-lines-free spectra to determine several properties of the stellar population, like metallicity, age, and star formation history. The error estimates of all these properties are also obtained. The weak spectral line features are then analyzed, and parameters are estimated without fitting the Gaussian profile. After the separate analysis of emission lines and stellar continuum is complete, their properties are recovered for each spaxel. The MaNGA Pipe3D value-added catalog has spatially resolved and integrated properties of galaxies for DR17. We have used some of the integrated properties of the galaxies like $\lambda_{Re}$, effective radius, mass-weighted age, and metallicity, flux intensity of H$\alpha$ line at 1 $R_e$, D4000 stellar index at 1 $R_e$, and inclination (extracted from the NASA Sloan Atlas\footnote{\url{http://nsatlas.org}}) We have also used the resolved maps of H$\alpha$ gas velocity, stellar velocity, H$\alpha$ emission line Gaussian flux, and spectral index D4000 in this work. For the visualization of these 2D maps, we used MaNGA-MARVIN \footnote{\url{https://dr17.sdss.org/marvin/}} \citep{Cherinka19}. The resolved maps are provided as 3D data cubes where two axes represent the spaxels and the third axis represents these parameters.

\subsubsection{SFR and stellar mass} \label{sec:SFR}
We use the GALEX-SDSS-WISE Legacy Catalog \citep[GSWLC-A2;][]{Salimcatalog16, SalimSBcatalog18}, hereafter the Salim catalog, for obtaining stellar mass, SFR, and redshift. The Salim catalog uses the stellar population synthesis technique to model the spectral energy distribution (SED) using ultraviolet, optical, and infrared bands to obtain parameters such as stellar mass and SFR for each galaxy. Input data are taken from the SDSS, Galaxy Evolution Explorer (GALEX), and Wide-field Infrared Survey Explorer (WISE) all-sky surveys. Several previous studies on star-forming early-type galaxies \citep[e.g.][]{Schawinski09, Kaviraj09, George15} have used optical or UV colours as the proxy for star formation activity. \citet{Salim15} has shown that $NUV-r$ color is good at indicating ongoing star formation in galaxies. Nevertheless, these color-based identifications of star-forming galaxies do not properly consider the effect of dust attenuation, which varies with the SFR and $M_{\ast}$ of a galaxy \citep[e.g.][]{Salim2020} and can lead to the wrong identification of a dusty star-forming galaxy as a passive, red galaxy \citep[e.g.][]{Cortese2012}. Determining SFR by modeling the SED is more robust than color-based techniques to identify star-forming galaxies. The Salim catalog allows us to select starforming or quenched galaxies using simple criteria based on the specific star formation rates (sSFR; the ratio of SFR and $M_{\ast}$) of galaxies determined from stellar population synthesis modeling. To select star-forming galaxies, we use 
\begin{equation} \label{eq:sf_selection}
    \rm{sSFR} (\log(\rm{yr}^{-1})) \geq -10.8 
\end{equation}
and
\begin{equation} \label{eq:q_selection}
    \rm{sSFR} (\log(\rm{yr}^{-1})) \leq -11.8 
\end{equation}
for quenched galaxies following \citet{Salim15}. Objects having $\rm{sSFR}(\log(\rm{yr}^{-1}))$ in between $-10.8$ and $-11.8$ are called green valley galaxies, which are in transition between the star-forming main sequence and quenched sequence \citep[see, e.g.][]{Bait17}.

\subsubsection{Environmental density and the $B/T$ 
 luminosity ratio} \label{sec:Environmental Density}
We use the local surface density of galaxies as given in \citet{Baldry06} to measure the environmental density. The density is defined as $\Sigma_N=N/\pi d_N^2$, where $d_N$ is the $N$th nearest neighbour distance, and $N$ is the number of galaxies inside the area $\pi d_N^2
$. This is calculated within the redshift range of $\pm \Delta zc=1000$ km s$^{-1}$ for galaxies where the spectroscopic redshift is known, or the photometric redshift is known within the 95\% confidence limit. The average $\Sigma_N$ for the fourth and fifth nearest neighbor is the final environmental density following \citet{Baldry06}.

We take the bulge-to-total ($B/T$) luminosity ratio from the MaNGA PyMorph \citep{Vikram10} DR17 photometric catalog, which provides photometric parameters of a galaxy from S\'{e}rsic and S\'{e}rsic+Exponential fits \citep{Fischer19, Dominguez22}. It is obtained by fitting the 2D surface brightness profiles of the SDSS-MaNGA DR17 final galaxy sample with a S\'{e}rsic+Exponential profile. 

\subsubsection{Morphology} \label{sec: Morphology}
For the morphological classification, we use the MaNGA morphology deep-learning DR17 catalog \citep[hereafter, DL17;][]{ Dominguez18, Fischer19, Dominguez21}. This is a deep-learning model-based classification of the final SDSS-MaNGA galaxy sample. The conventional neural network algorithm is trained on two catalogs: \cite{Nair10}, which is a visual morphology classification of the SDSS DR7 \citep{Abazajian09} images, and Galaxy Zoo 2 \citep{Willett13}. It claims an accuracy greater than 90\% on the deep-learning classification. Based on the T-Type, this catalog separates the late and early-type galaxies better than the Galaxy Zoo MaNGA catalog. It also gives finer separation between pure ellipticals and S0s. All the deep-learning-based classifications were eye-balled for additional reliability, and the catalog provides a visual classification (VC) number and its reliability flag (VF). Here, we only choose galaxies with certain visual classification (VF $=0$). For a clean separation of the galaxies in different morphology classes, \citet{Dominguez21} recommend:
\begin{itemize}
    \item For LTGs: T-Type $> 0$ AND P\_LTG $\geq 0.5$ AND VC $=3$
     \item For S0s: T-Type $\leq 0$ AND P\_S0 $> 0.5$ AND P\_LTG $< 0.5$ AND VC $=2$
    \item For Es: T-Type $\leq 0$ AND P\_S0 $\leq 0.5$ AND P\_LTG $< 0.5$ AND VC $=1$
\end{itemize}
P\_LTG is the probability of a galaxy being LTG rather than ETG, and P\_S0 is the probability of a galaxy being S0 rather than a pure elliptical. We follow this prescription to assign a morphology to the galaxies in our sample.

\subsection{Sample selection} \label{sec:Sample selection}
For our analysis, we cross-matched all the catalogs for the mentioned quantities and used appropriate flags mentioned in the catalogs to select only galaxies not flagged in any catalog. For GSWLC, we used the SED fitting flag ($flag\_sed = 0$) and reduced goodness-of-fit value for the SED fitting ($\chi^2_r \geq 5$); for DL17 we used the Visual classification flag ($Visual\_Flag = 0$) and for PyMorph we used S\'{e}rsic fit flag ($FLAG\_FAILED\_S = 0$) and S\'{e}rsic+Exponential fit flag ($FLAG\_FAILED\_SE = 0$). We also checked if all the galaxies in the sample have Pipe3D outputs. As we are interested in nearby galaxies, we put a cut on the redshift of $z<0.1$. All these selection criteria produce 5076 galaxies. We tested for any gross biases produced by these selections. The distribution of stellar mass, sSFR, and morphology before and after applying the criteria are very similar. So, we believe these selection cuts did not produce any large biases beyond those already present in the MaNGA selection.

\begin{figure*}
    \centering
    \includegraphics[width = 1\textwidth]{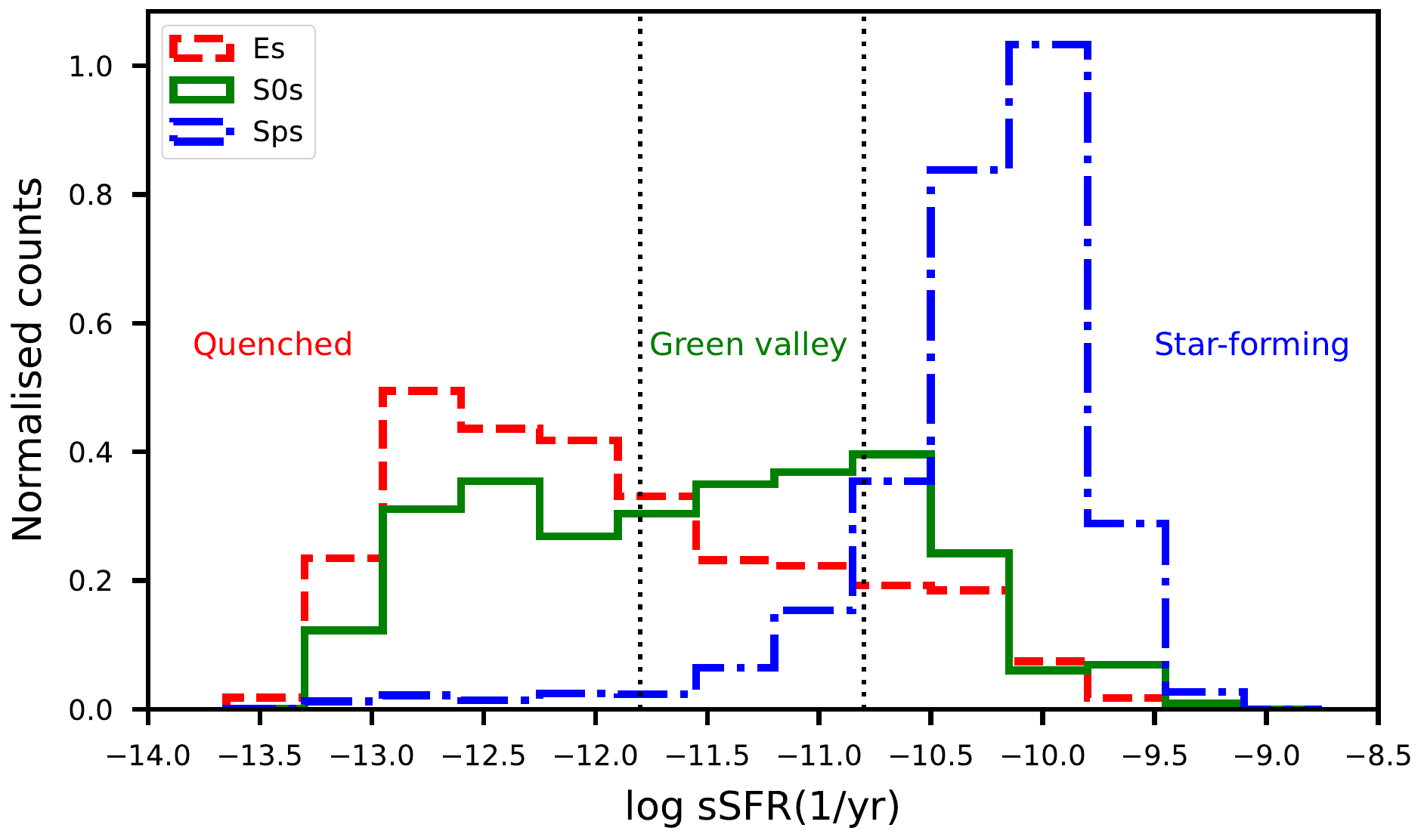}
    \caption{The normalized counts (area under the weighted histogram is one) of galaxies of different morphological types as a function of log(sSFR(1/yr)). A total of 3099 spiral galaxies are shown by a blue dashed-dotted line, a green solid line shows 607 S0s, and 1139 elliptical galaxies are shown by a red dashed line. By dotted black lines ($\rm{sSFR} (\log(\rm{yr}^{-1})) = -10.8$ and $\rm{sSFR} (\log(\rm{yr}^{-1})) = -11.8$), the border between the quenched, green valley, and star-forming regions are also shown. Almost all the Sps are in the star-forming region. S0s are equally distributed in quenched and green valley regions and decrease in the star-forming region. Es are mostly quenched, and their population decreases as the sSFR increases but not as rapidly as the population of Sps decreases with decreasing sSFR. The weighted median value of the log(sSFR(1/yr)) for Sps, S0s, and Es are  $-10.17$, $-11.47$, and $-12.03$, respectively. Furthermore, we can also see that a significant fraction of S0s and Es are star-forming. The Es, which are in the star-forming region, form our primary sample.}
    \label{fig: ssfr}
\end{figure*}

First, we divide our whole sample into three morphological classes: ellipticals (Es), lenticulars (S0s), and spirals (Sps). We used the criteria discussed in Section \ref{sec: Morphology} to split the whole sample into these three samples.  After this division, we have 1139 Es, 607 S0s, and 3108 Sps. Fig. \ref{fig: ssfr} shows the distribution of the galaxies in sSFR. We can see that spiral galaxies are in the star-forming region, and ellipticals are mostly quenched. In this work, we are interested in elliptical galaxies in the star-forming region. We can also notice that a significant fraction of S0s are also star forming. Like ellipticals, S0 galaxies are also known to be non star forming and red \citep[e.g.][]{Martig09, Bait17, Mishra19}. So, the study of star forming S0s can also give important insights into the evolution of early-type galaxies \citep[e.g.,][]{Rathore22, Xu21}.

We then subdivided our sample based on the sSFR and morphology into three subsamples:
\begin{itemize}
    \item Star-forming ellipticals (SF-Es; elliptical galaxies with $\rm{sSFR} (\log(\rm{yr}^{-1})) \geq -10.8$)
    \item Star-forming spirals (SF-Sps; spiral galaxies with $\rm{sSFR} (\log(\rm{yr}^{-1})) \geq -10.8$)
    \item Quenched ellipticals (Q-Es; ellipticals with $\rm{sSFR} (\log(\rm{yr}^{-1})) \leq -11.8$).
\end{itemize}
The first is our main sample of interest, the Primary sample and the latter two are control samples, which are used to compare different quantities across these samples and draw conclusions from them. The choice of these three samples is motivated by the scientific problem. We are studying the evolution of star-forming ellipticals and how they form stars, unlike their regular counterparts. Also, we aim to investigate if there is any evolutionary connection present between these samples. Stellar mass is a fundamental quantity and plays an essential role in galaxy evolution \citep{Peng2010}. For this reason, we restrict the stellar mass of the control samples (SF-Sps and Q-Es) to the same range as that of the primary sample (SF-Es).

\subsubsection{Sample of Star-forming ellipticals (SF-Es)} \label{sec: SF-E}
After the basic selection cuts, we use the DL17 criteria and Equation \ref{eq:sf_selection} to identify elliptical galaxies that are star forming. Using these conditions, we obtained 76 star-forming elliptical galaxies.

To identify any morphological misclassification done by the DL17, we carry out a careful visual inspection of the SF-Es in the deeper imaging of Dark Energy Spectroscopic Instrument (DESI) Legacy Survey DR10 \citep{Schlegel2021} using the Legacy survey Sky Browser\footnote{\url{https://www.legacysurvey.org/viewer/}}. See Appendix \ref{sec: imgs}, Figure \ref{fig:fig_A4}, for all the images of the final sample of SF-Es. For all the SF-Es, Legacy DR10 images were available. We discarded two galaxies showing spiral features in DR10 images (see Appendix \ref{sec: imgs}, Figure \ref{fig:fig_A1}), which likely represent a failure of the DL17 classification. We also looked carefully at edge-on galaxies as their morphological classification is tricky because it is possible to misclassify an edge-on spiral galaxy with a large bulge as an S0 galaxy. This happens because spiral features of spiral galaxies are often invisible at large inclination angles.  We did not find any high inclination galaxy in our sample. If the disks of S0s lack bright stars, they tend to be smooth, which makes it challenging to separate S0s from ellipticals,  particularly for face-on galaxies. Due to this effect, the sample could contain some S0 galaxies.

Further, we check the H$\alpha$ emission line maps and the GALEX data for all the SF-Es. We checked to see if the sSFR value was incorrectly computed due to the presence of a nearby star-forming galaxy. We discarded 14 galaxies without a trace of H$\alpha$ emission in their emission maps. These galaxies were also not detected in GALEX. GALEX detection is a valuable diagnostic because UV luminosity is a measure of the recent star formation activity in the galaxy. We also discarded one galaxy that does not have Pipe3D output. Finally, we have 59 SF-Es left in the sample. The redshift and stellar mass ranges of the 59 SF-Es are respectively $0.01 < z < 0.08$ and $8.58 < \log \frac{M_\ast}{M_{\odot}} < 11.31$. In Appendix \ref{sec: table}, Table \ref{tab: SFE_Table}, we represent the different parameters of the first 10 galaxies from the primary sample of SF-Es.

\subsubsection{Control Sample Of Star-forming Spirals (SF-Sps)}
\label{sec: SF-Sp}
This is our first control sample of star-forming spirals. After applying the basic selection criteria, we select spirals based on the DL17 criteria (for LTG) and use Equation \ref{eq:sf_selection} to identify star-forming galaxies. We have also made an additional cut to match the stellar mass range of the SF-Sps with that of the SF-Es. These conditions, taken together, yield 2419 star-forming spiral galaxies. These 2419 galaxies make up our control sample of SF-Sps. Given the large sample size, we did not visually inspect these galaxies. Also, it would be difficult to misclassify any other morphology to spiral. Considering the claimed accuracy of DL17, there could be at most 10\% contamination in this sample. Appendix \ref{sec: table}, Table \ref{tab:SFSp_Table} represents various parameters for the first 10 galaxies from this control sample of SF-Sps.

\subsubsection{Control Sample Of Quenched ellipticals (Q-Es)}
\label{sec: Q-Es}
To select ellipticals, we use the DL17 criteria as the primary sample of SF-Es and equation \ref{eq:q_selection} to separate the quenched ellipticals. Further, we again make a selection cut to match the stellar mass range of the sample to be the same as that of the SF-Es.  After the basic selection cuts and filtering according to these metrics, we have 684 quenched ellipticals. These 684 galaxies are part of the second control sample of Q-Es. For this sample, we also did not carry out a visual inspection of morphology. Considering the accuracy of DL17 and any other morphological class of galaxies to be truly passive, like ellipticals, there could be contamination from some quenched S0s in this control sample. The various parameters for the first 10 galaxies from the control sample of Q-Es are represented in Appendix \ref{sec: table}, Table \ref{tab:QE_Table}.

\begin{figure*}
    \centering
    \includegraphics[width = 0.325\textwidth]{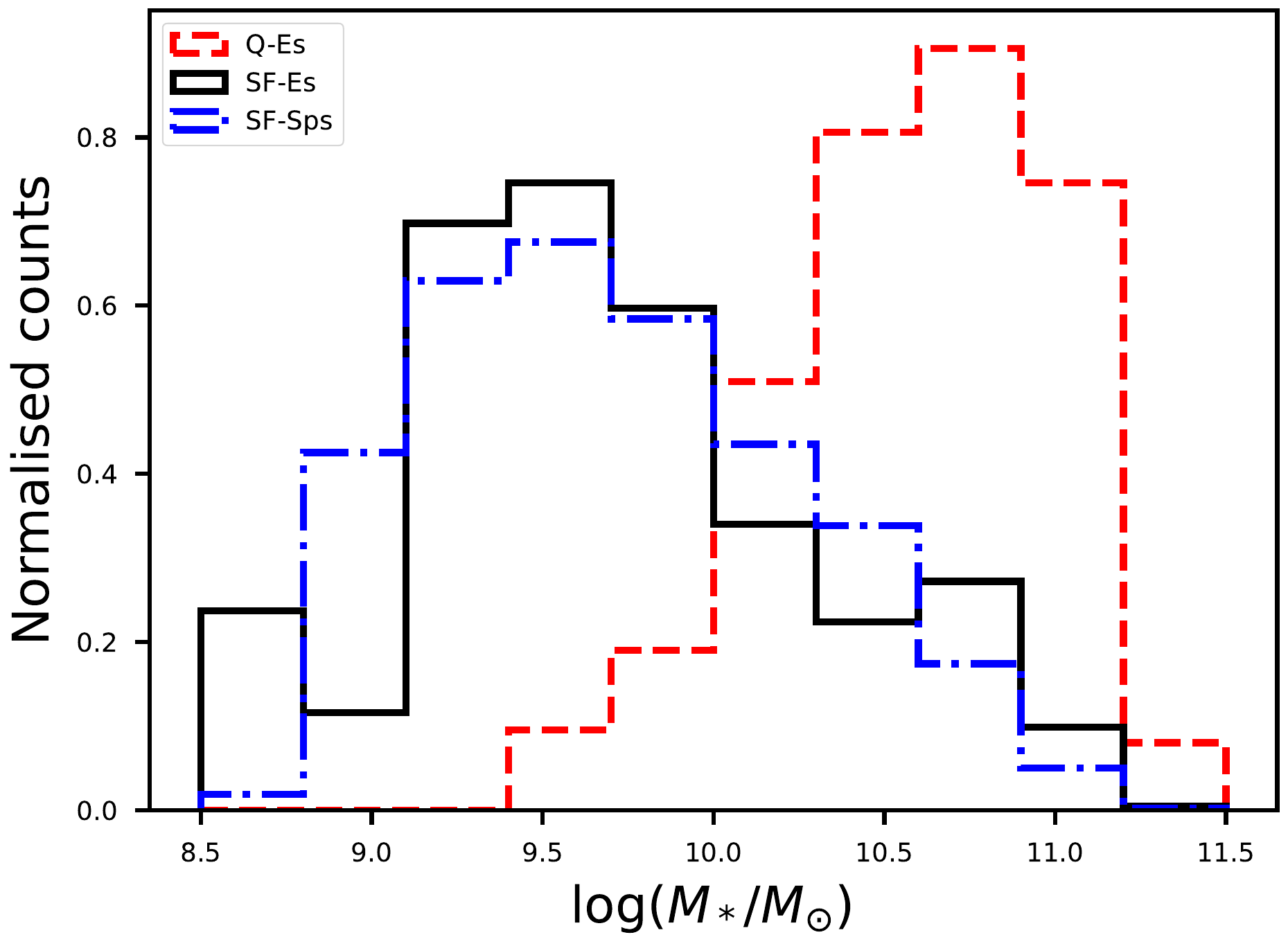}
    \includegraphics[width = 0.325\textwidth]{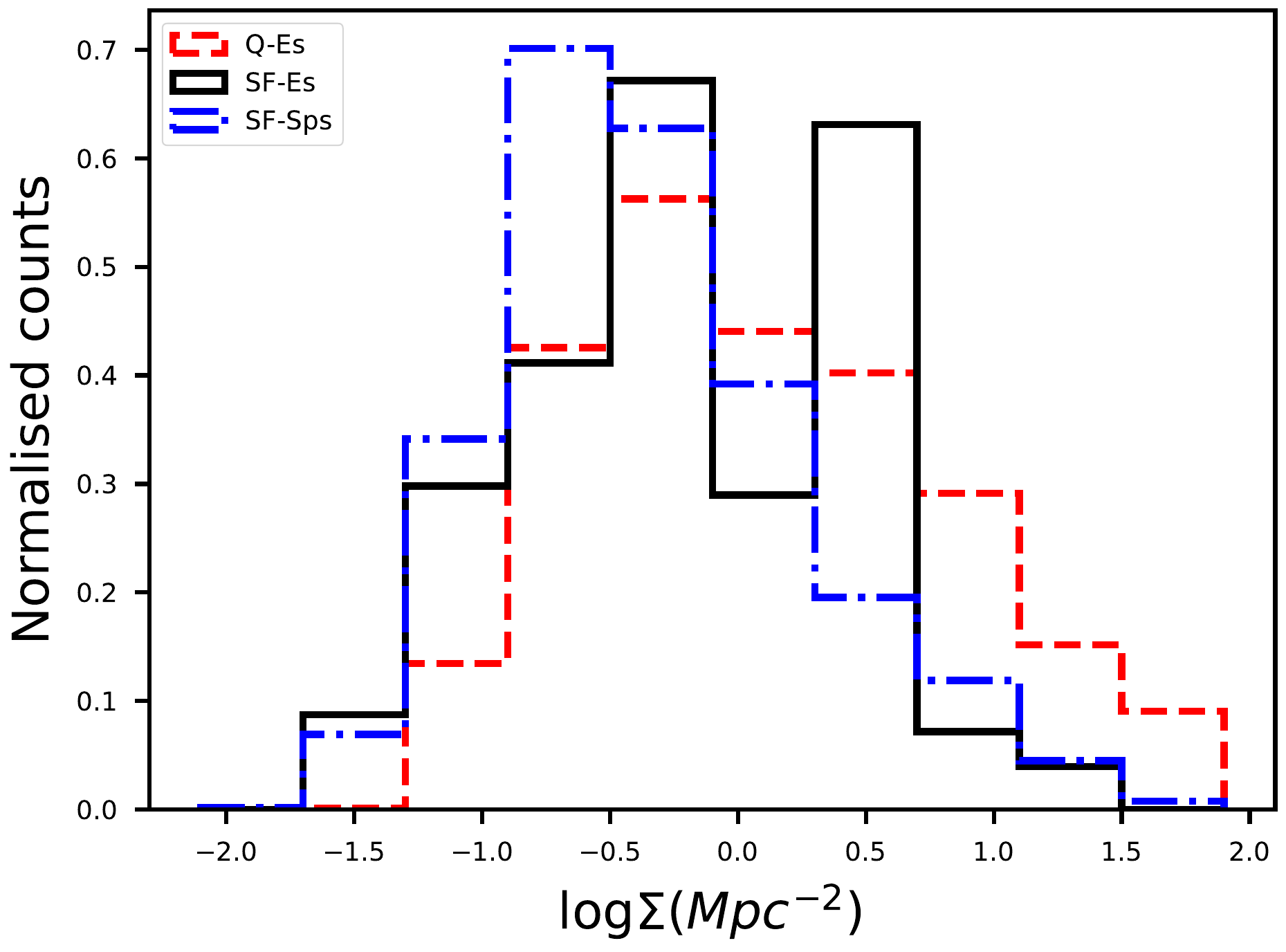}
    \includegraphics[width = 0.325\textwidth]{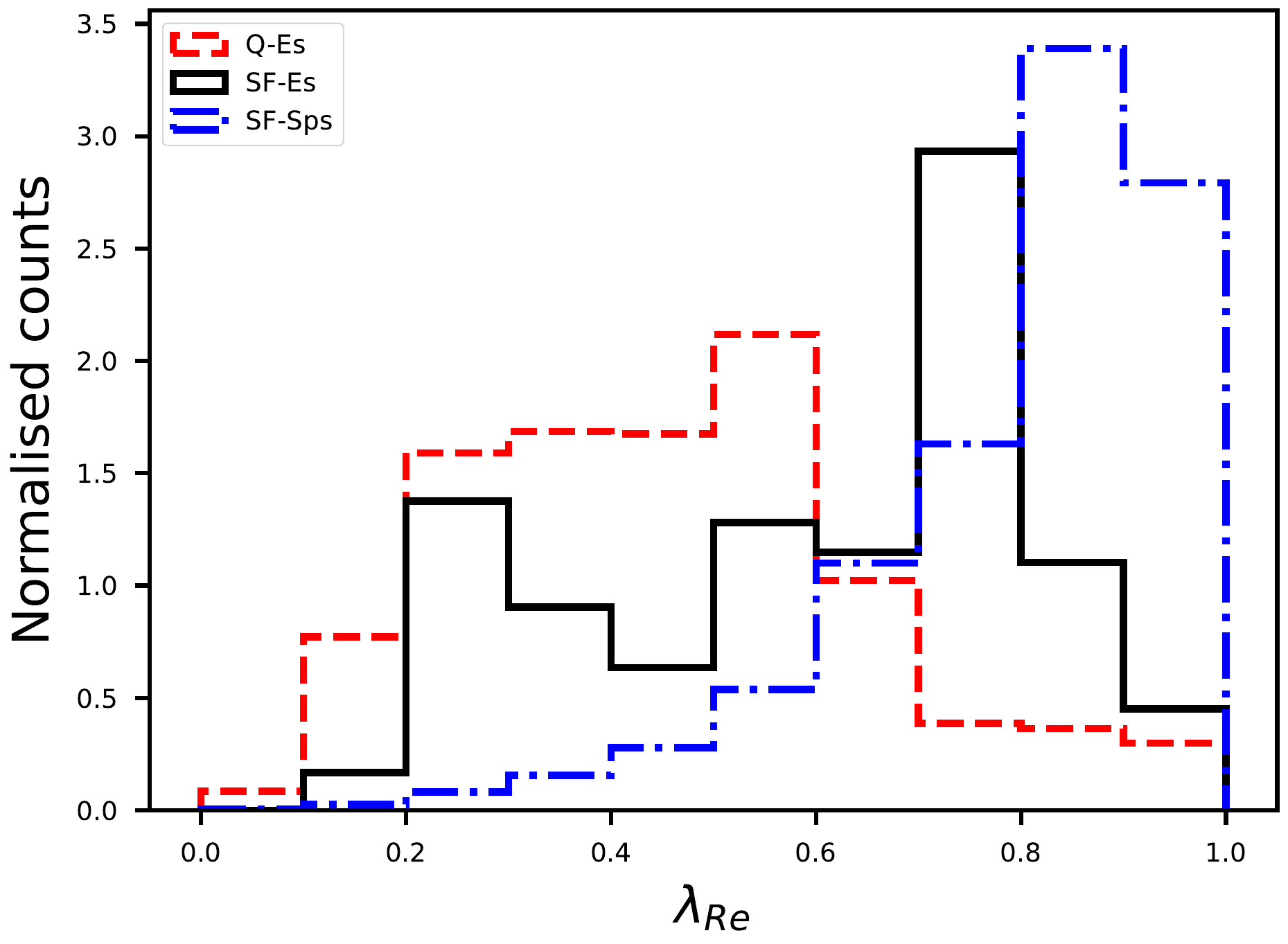}
    \caption{ For all the panels, 2419 SF-Sps are shown by a blue dashed-dotted line, 59 SF-Es are shown by a black solid line, and by a red dashed line, 684 Q-Es are shown. Left panel: The normalized counts of galaxies of the three samples as a function of stellar mass (log$\frac{\rm{M}_{*}}{\rm{M}_{\odot}}$). The stellar mass distribution of the SF-Es and SF-Sps are similar and have the same weighted median stellar mass of $10^{9.67}$ M$_{\odot}$.  Also, on average, SF-Es and SF-Sps are less massive than the Q-Es, which have a weighted median stellar mass of $10^{10.62}$ M$_{\odot}$.  Middle panel: The normalized counts of galaxies of the three samples as a function of environmental density (log$\Sigma (\rm{Mpc}^{-2})$). The median values of log$\Sigma (\rm{Mpc}^{-2})$ for SF-Sps, SF-Es, and Q-Es are $-0.43$, $-0.23$, and $0.01$, respectively. The distribution of galaxies across the three samples as a function of environmental density is more or less similar. However, on average, the value of $\Sigma (\rm{Mpc}^{-2})$ increases from SF-Sps to SF-Es and SF-Es to Q-Es. Right panel: The normalized counts of galaxies of the three samples as a function of $\lambda_{Re}$.  Q-Es are generally slow rotators, and SF-Sps are fast rotators. However, the SF-Es have a mixture of slow and fast rotators. The median value of $\lambda_{Re}$ for SF-Sps, SF-Es, and Q-Es are $0.84$, $0.66$, and $0.46$ respectively.}.
    \label{fig: massenv}
\end{figure*}

\begin{figure}
    \centering
    \includegraphics[width = 0.5\textwidth]{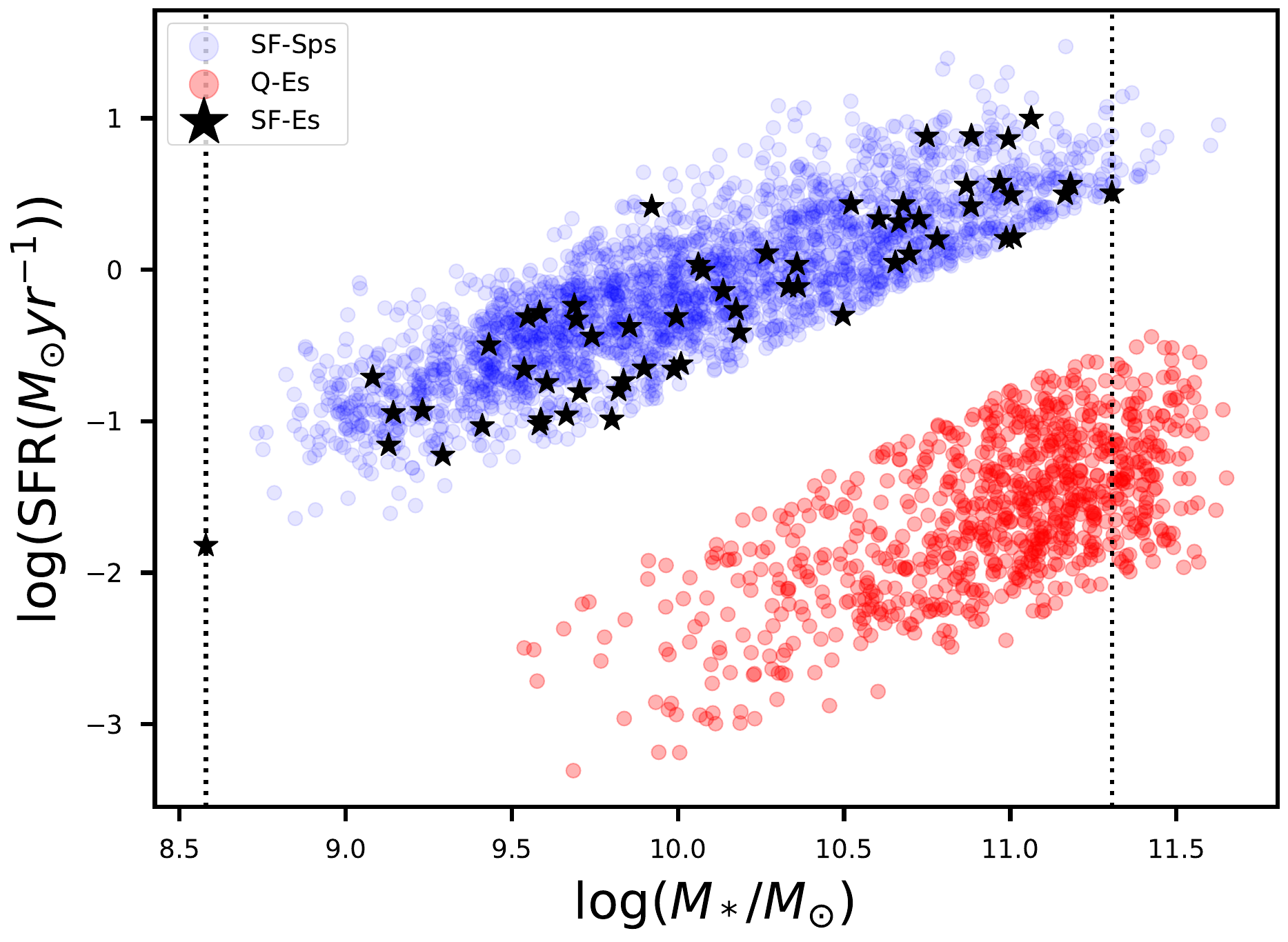}
    \caption{Distribution of the galaxies across the three samples in the log(SFR(M$_{\odot} \rm{yr}^{-1}))$ and log($\frac{\rm{M}_{\ast}}{\rm{M}_{\odot}}$) plane (star-forming main sequence). Blue circles show 2436 star-forming spirals, black stars show 59 star-forming ellipticals and 839 quenched ellipticals are shown by red circles. The black dotted lines ($\rm{log}(\frac{\rm{M}_{\ast}}{\rm{M}_{\odot}})=8.58$ and $\rm{log}(\frac{\rm{M}_{\ast}}{\rm{M}_{\odot}})=11.31$) show the stellar mass range of the primary sample of SF-Es. We matched the stellar mass ranges of the control samples of Q-Es and SF-Sps to the primary sample by selecting galaxies within the two dotted lines. The galaxies outside the lines are discarded from further analysis. This selection results in 2419 SF-SPs and 684 Q-Es, constituting the two final control samples. The weighted median values of log(SFR(M$_{\odot} \rm{yr}^{-1}))$ for SF-Sps, SF-Es, and Q-Es are $-0.43$, $-0.66$, and $-1.92$, respectively.}
    \label{fig: sfms}
\end{figure}

\subsection{Global Properties} \label{sec: global props}
Fig. \ref{fig: massenv} (left panel) shows the stellar mass distribution for SF-Es, SF-Sps, and Q-Es. The star-forming spirals generally have lower stellar mass than the quenched ellipticals. This is consistent with the fact that, on average, star-forming LTGs are less massive, and quiescent ETGs are more massive \citep{Baldry04, Con06, Buitrago13, Bait17}. Surprisingly, the primary sample of SF-Es has a stellar mass distribution similar to SF-Sps. Also, the weighted median of the stellar mass for SF-Es and SF-Sps is the same, $10^{9.67}$ M$_{\odot}$, and that for Q-Es is significantly higher at  $10^{10.62}$ M$_{\odot}$. The population of SF-Es and SF-Sps decreases; on the other hand, the population of Q-Es increases rapidly beyond the stellar mass of  $10^{10}$ M$_{\odot}$. Most SF-Es have a stellar mass of less than $10^{10.25}$ M$_{\odot}$, as seen in most previous studies \citep{Huertas10, Kannappan09} involving blue or star-forming ETGs. However, with increasing stellar mass, the number of SF-Es does not decrease as rapidly as has been seen with only star-forming S0s \citep{Rathore22}. This could be because elliptical galaxies are generally more massive than S0s. Here, we can see that the number of lower stellar mass Q-Es is very few. Quenched galaxies tend to be more massive, particularly ellipticals. However, selection bias against Q-Es at lower masses could be present due to limitations of stellar population modeling. These less massive quenched galaxies have very low luminosity, particularly for the bluer bands, resulting in low flux for a given redshift. This makes it difficult or impossible to obtain reliable stellar population estimates for these galaxies from the SPS modeling using UV-Optical-mid-IR band photometry.

Fig. \ref{fig: massenv} (middle panel) shows the environmental-density-weighted histogram for SF-Es, SF-Sps, and Q-Es.  Here, the average $\Sigma_N$ for the fourth and fifth nearest neighbor is taken as the environmental density.  We can see that the overall distribution of the galaxies with $\Sigma$ is similar. The number of Q-Es is highest for higher $\Sigma$ bins, and SF-Sps dominate in the lower $\Sigma$ bins. This is consistent with quenched ETGs residing in denser regions \citep{Dressler1980, Bait17}. The primary sample of SF-Es lies between the SF-Sps and Q-Es. The weighted median values of log$\Sigma (\rm{Mpc}^{-2})$ for SF-Sps, SF-Es, and Q-Es are $-0.43$, $-0.23$, and $0.01$, respectively.

Fig. \ref{fig: massenv} (right panel) shows the $\lambda_{Re}$ distribution of galaxies for the three samples. SF-Sps are, on average, fast rotators and dominate the two highest bins in $\lambda_{Re}$.  Q-Es are slow rotators whose population decreases quickly after $\lambda_{Re} = 0.6$. However, the primary sample of SF-Es is a mixture of fast and slow rotators.   The weighted median values of $\lambda_{Re}$ for SF-Sps, SF-Es, and Q-Es are $0.84$, $0.66$, and $0.46$, respectively.

Actively star-forming galaxies lie on a relatively tight sequence in the SFR stellar mass plane known as the star-forming main sequence (SFMS). Fig. \ref{fig: sfms} shows the SFMS for our samples. SF-Es are distributed in the SFMS along with the SF-Sps. The weighted median values of the SFR in units of M$_{\odot} yr^{-1}$ for the SF-Sps, SF-Es, and Q-Es are $-0.43$, $-0.66$, and $-1.91$, respectively. In this work, our goal is to understand this population of SF-Es, which lies in the SFMS and not in the quenched region, with the majority of ellipticals.

In the next section, we will describe the results and compare the three samples with each other to gain a clearer insight into the nature of the SF-E population.

\begin{figure*}
    \centering
    \includegraphics[width = 0.47\textwidth]{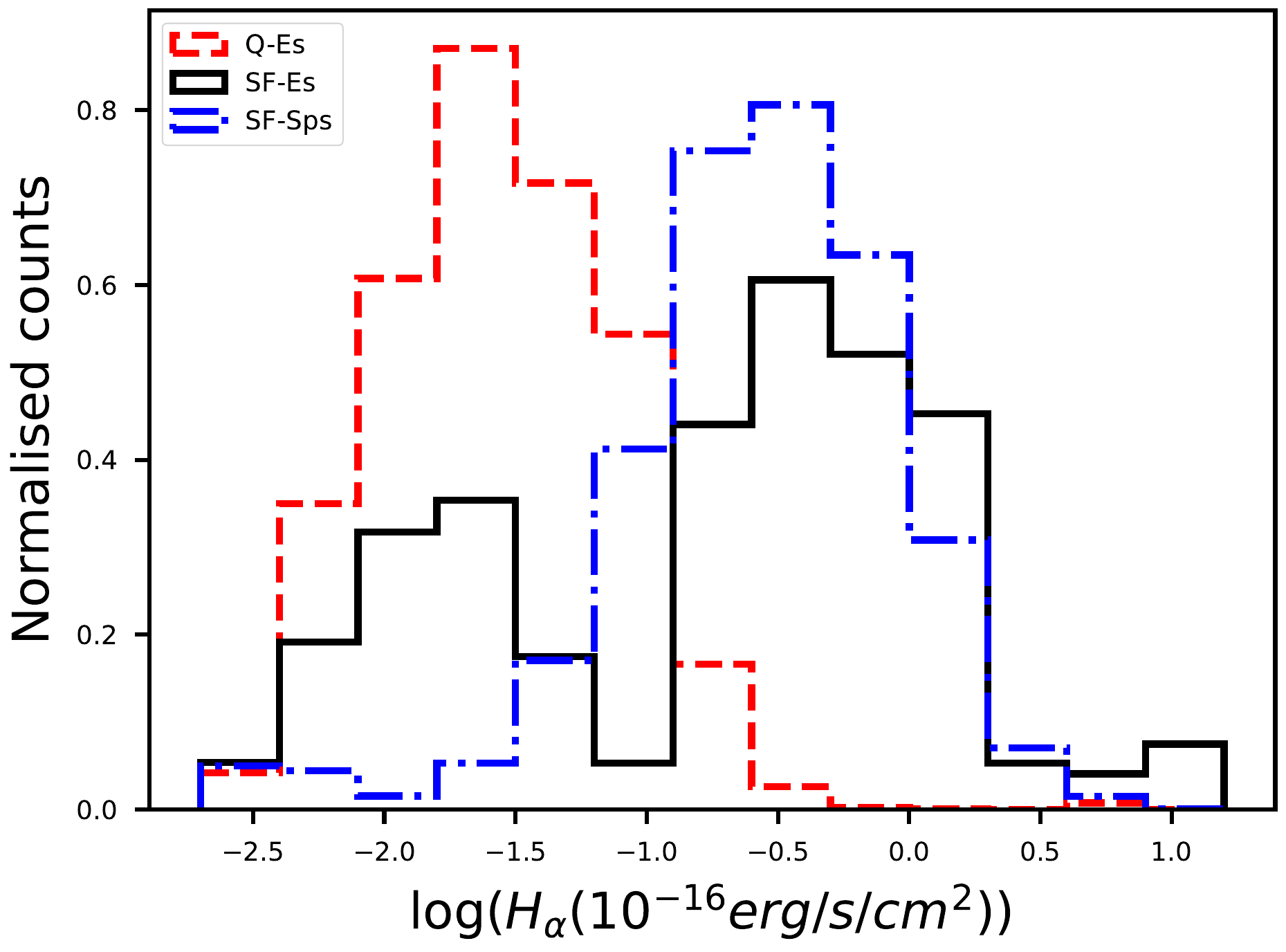}
    \includegraphics[width = 0.47\textwidth]{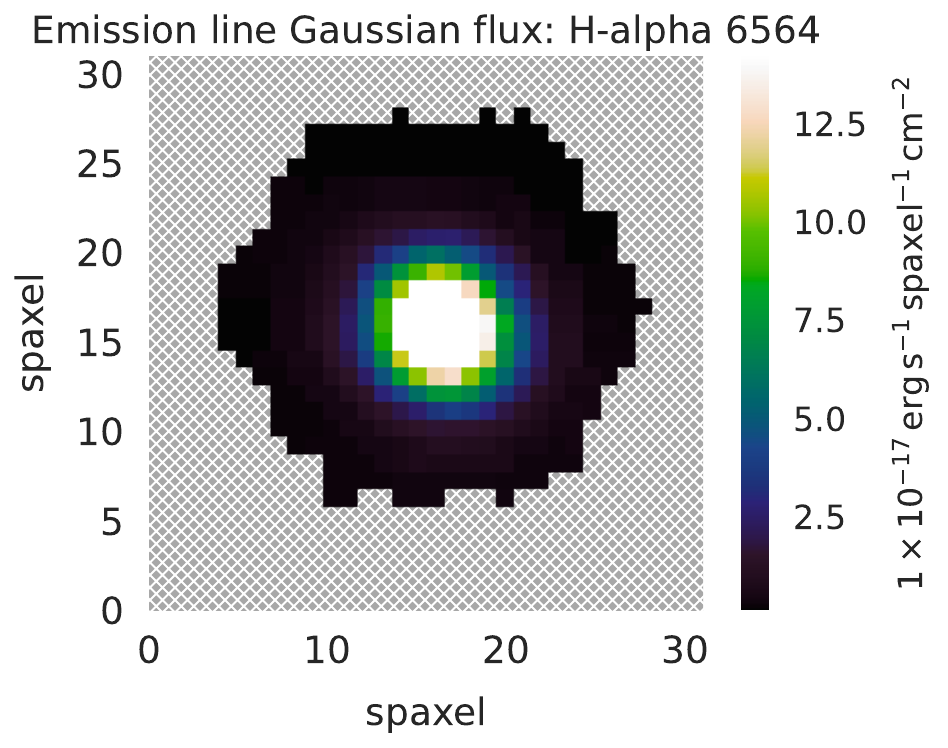}
    \caption{We have taken the \textbf{dust-corrected} H$\alpha$ emission line Gaussian flux as the measure of the gas content of the galaxy. This figure (left panel) shows the normalized counts (weighted histogram) of the three samples of galaxies (Q-Es in red, SF-Es in black, and SF-Sps in blue) as a function of $\log(\rm{H}_{\alpha}(10^{-16} \rm{erg/sec/cm}^2))$. We can see that the gas content of Q-Es is low compared to the SF-Sps. However, the primary sample of SF-Es shows two distinct populations. In one, the gas content is like SF-Sps; in the other, it is like Q-Es. The weighted median values of $\log(\rm{H}_{\alpha}(10^{-16} \rm{erg/sec/cm}^2))$ for SF-Sps, SF-Es, and Q-Es are $-0.54$, $-0.57$, and $-1.57$ respectively. The H$\alpha$ emission map of one of the SF-Es is shown in the right panel.}.
    \label{fig: h-alpha}
\end{figure*}

\begin{figure*}
    \centering
    \includegraphics[width = 0.47\textwidth]{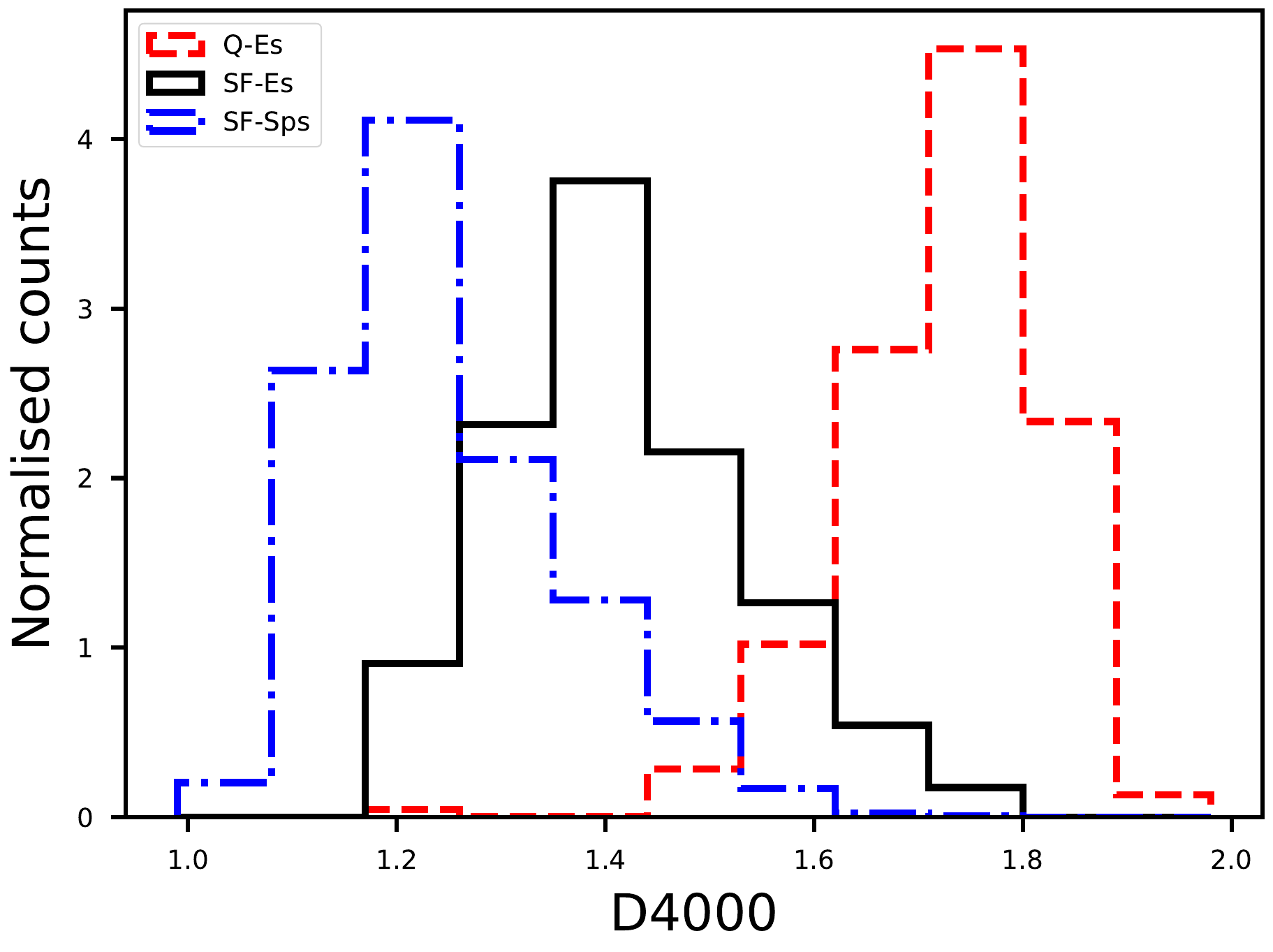}
    \includegraphics[width = 0.47\textwidth]{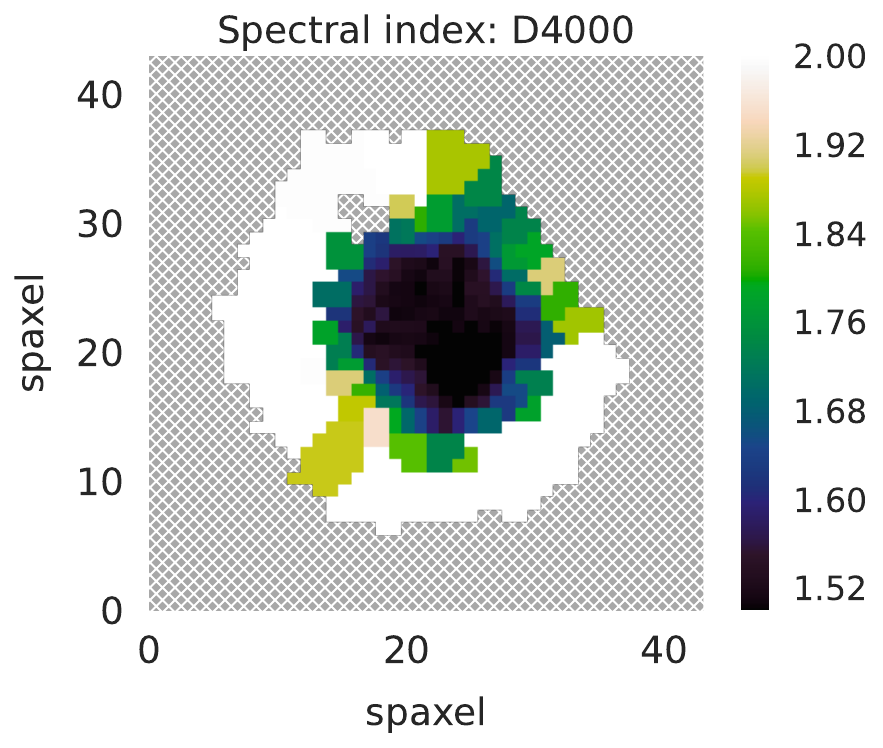}
    \includegraphics[width = 0.47\textwidth]{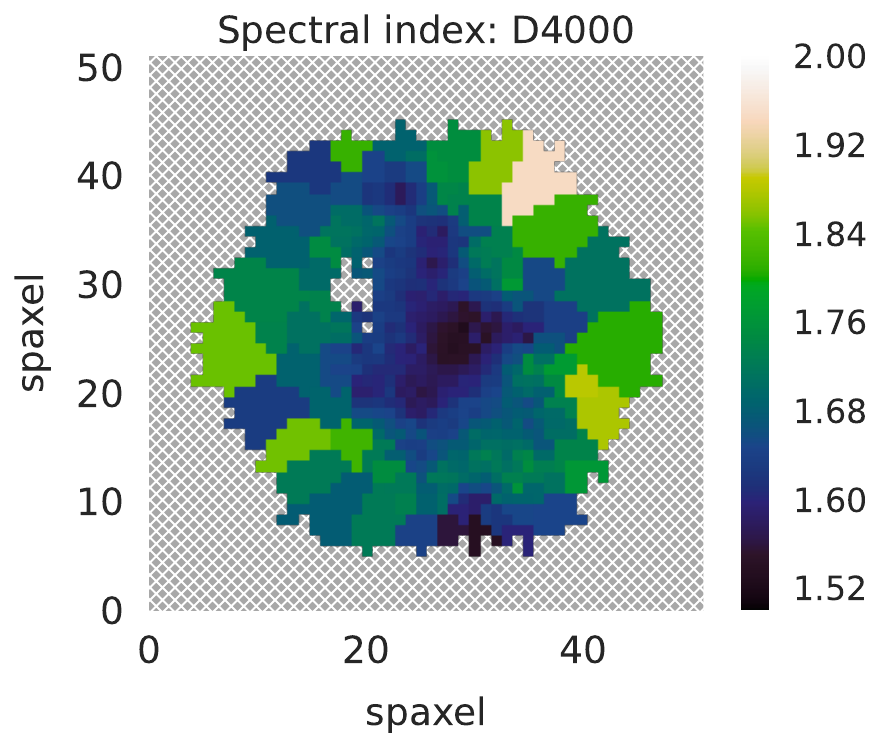}
    \includegraphics[width = 0.47\textwidth]{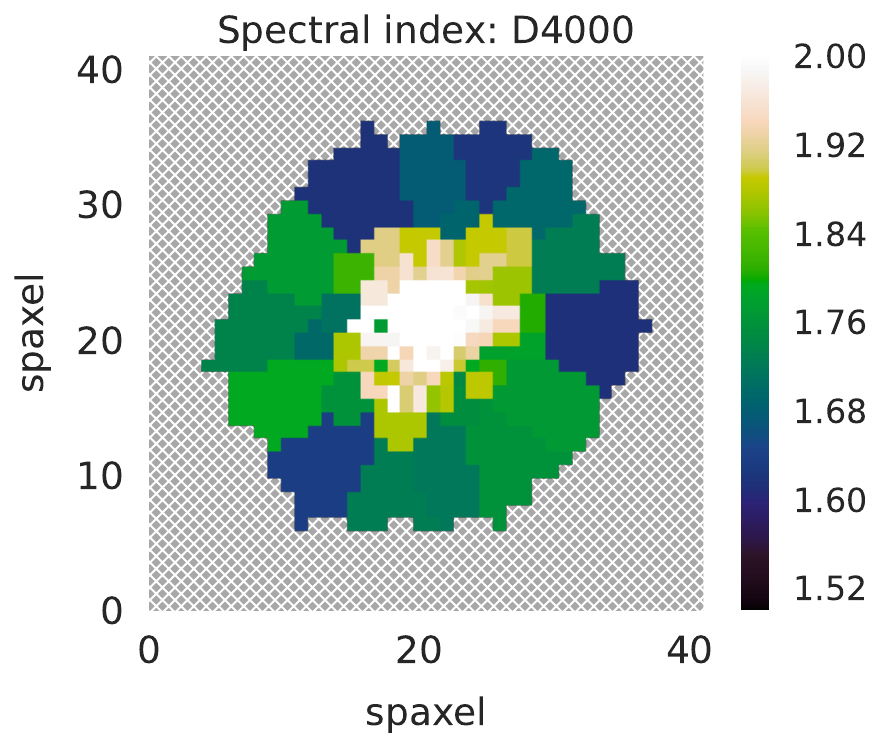}
    \caption{The upper left panel shows the normalized counts (weighted histogram) of the three samples of galaxies (Q-Es in red, SF-Es in black, and SF-Sps in blue) as a function of the D4000 spectral index, which is a continuum discontinuity at 4000 \AA. The value of the D4000 increases with the age of the stellar population and the abundance of metal in a galaxy. We can see that the two control samples of SF-Sps and Q-Es have young and old stellar populations, respectively. However, the primary sample of SF-Es has a mixture of old and young stellar populations. The weighted median values of D4000 for SF-Sps, SF-Es, and Q-Es are $1.23$, $1.41$, and $1.74$, respectively. We have also shown three types of distribution of the resolved D4000 index across the 2D face of three galaxies. Upper right panel: low D4000 at the center and high at edges. Lower right panel: high at the center and low at the edges. Lower left panel: the low value of D4000 across the galaxy.}.
    \label{fig: D4000}
\end{figure*}

\section{Results} \label{sec: results}
From Pipe3D, we obtain the integral and the resolved properties of individual galaxies. In this section, we present the findings about these properties of our primary sample of SF-Es in comparison with the two control samples. In particular, we compare, in turn,  the gas content, size, mass, age, metallicity, $B/T$, environmental density, and velocity maps of the galaxies. 

\subsection{Gas content} \label{sec: gas content}
We have taken the dust-corrected flux intensity of the H$\alpha$ line within one $R_e$ as a proxy for the ionized gas content of the galaxy. H II regions are usually associated with star formation and are detected through their H$\alpha$ emission. As shown in Fig. \ref{fig: h-alpha} (left panel), H$\alpha$ emission from SF-Sps and Q-Es are well separated. This is as expected; the Q-Es have low gas content, while the SF-Sps have high gas content. There are two populations of galaxies in the SF-Es; one has low gas content like the Q-Es, and the other has gas as high as the SF-Sps. From pipe3D, we also have maps of the dust-corrected emission line gas flux of H$\alpha$. In our primary sample, almost all of the SF-Es show an H$\alpha$ blob in the center of the map (see Fig. \ref{fig: h-alpha} right panel). In Appendix \ref{sec: imgs}, Figure \ref{fig:fig_A3}, we presented H$\alpha$ maps for all the SF-Es.

\subsection{Spectral index: D4000} \label{sec: d4000}
Spectra of early-type galaxies show a red continuum, absorption lines due to the evolved older population of stars and little or weak emission lines because of the lack of (or very weak) star formation. They also show a continuum discontinuity at 4000 \text{\AA}. This results from the accumulation of absorption lines due to ionized metals in the atmosphere of old stars and the absence of hot, blue stars in the galaxy. So, the strength of the 4000 \text{\AA} break increases with stellar age (elapsed time since the last star formation activity) and metallicity. The value of this continuum break is measured as the spectral index and calculated as a logarithm of the ratio of integrated flux density redward and blueward of 4000 \text{\AA}. Moreover, the value of this spectral index can be used to separate old (D4000 $\sim 1.6-2.0)$ and young (D4000 $\sim 1.1-1.4)$ stellar populations.

Fig. \ref{fig: D4000} (upper left panel) shows that the SF-Sps (D4000 $\sim 1.1-1.5)$ have a young stellar population in contrast to the Q-Es (D4000 $\sim 1.6-1.9)$, which have an old stellar population. The value of the spectral index D4000 varies between $1.2$ and $1.6$ for the SF-Es. So, our sample of SF-Es has a mixture of young and old stellar populations or a population of intermediate-age stars. We also have the spectral index maps for all of these galaxies. Most of the D4000 maps of SF-Es are like the one shown in Fig. \ref{fig: D4000} (upper right panel), with a young population of stars in the center of the galaxy and old stars toward the outer side. We also have some maps showing young stellar populations in the outer regions and older stellar populations in the center (Fig. \ref{fig: D4000} lower right panel). Very few of them show a young stellar population distributed all over the galaxy, as shown in Fig. \ref{fig: D4000} (lower left panel). Spectral index D4000 maps of all the SF-Es are shown in \ref{sec: imgs}, Figure \ref{fig:fig_A3}.

\subsection{Age and metallicity} \label{sec: age-metallicity}
\begin{figure}
    \centering
    \includegraphics[width = 0.5\textwidth]{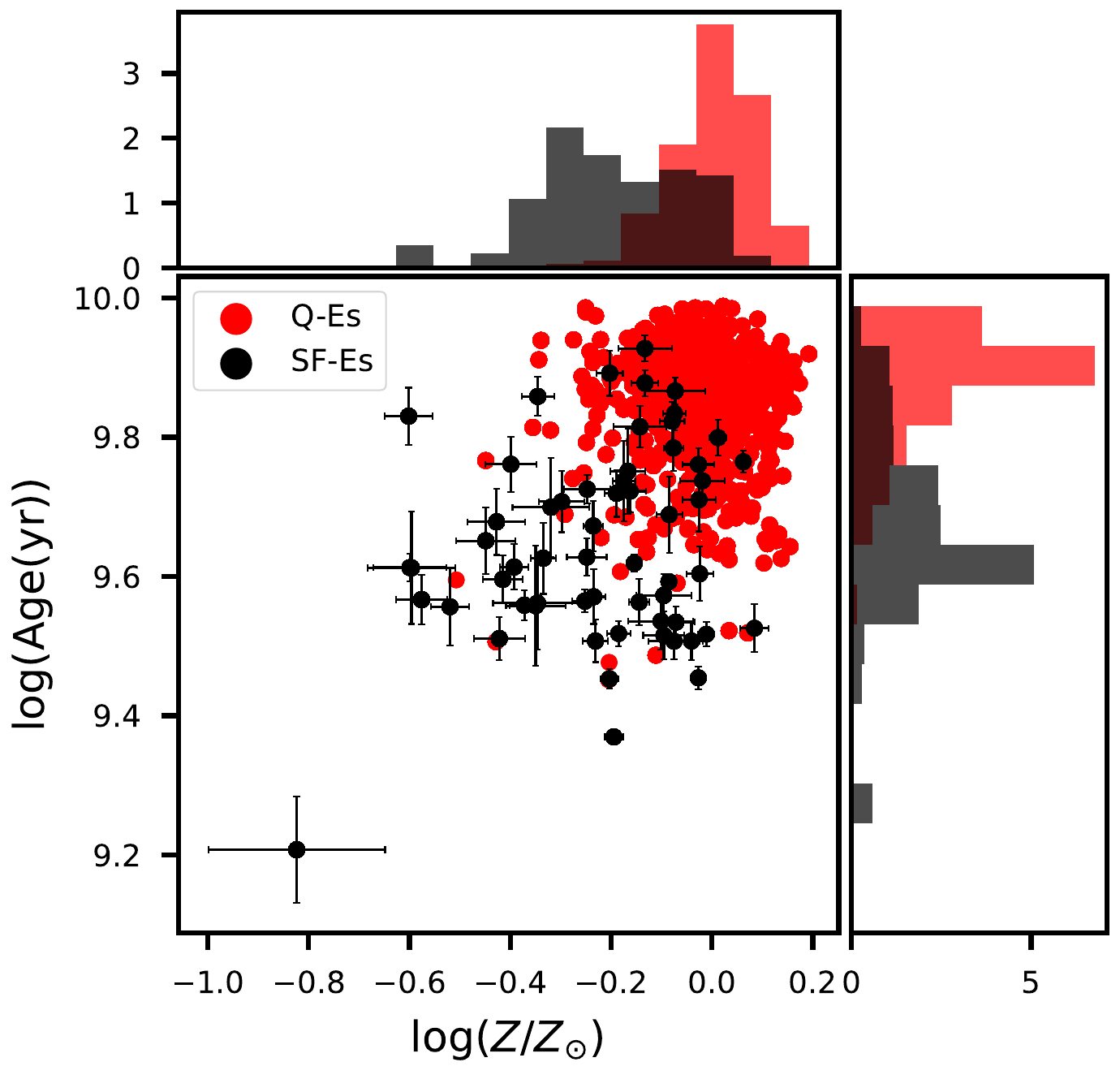}
    \caption{Distribution of the two samples of ellipticals SF-Es and Q-Es in the mass-weighted stellar population age-metallicity plane. Both the age and metallicity of the stellar population are, on average, smaller for the SF-Es compared to the Q-Es, suggesting that the SF-Es have acquired metal-poor gas in the recent past. The weighted median values of $\log(age(yr))$ for SF-Es and Q-Es are $9.61$ and $9.86$ respectively. The weighted median values of $\log(z/z_{\odot})$ for SF-Es and Q-Es are $-0.35$ and $-0.04$, respectively. Here, we can see that the histograms and the scatter plot are not exactly in agreement because the histograms are weight corrected, but the scatter plot is not. We do not correct the scatter plots since the values of the parameters are correct at the individual galaxy level.}.
    \label{fig: age_metallicity}
\end{figure}

We plotted two elliptical samples (SF-Es and Q-Es) on the age metallicity plane in Fig. \ref{fig: age_metallicity}. Here, age and metallicity are the mass-weighted age and metallicity of the stellar population. As we can see from Fig. \ref{fig: age_metallicity}, both age and metallicity of the stellar population for SF-Es are lower, on average, than the Q-Es. This suggests that, in the recent past, these SF-Es must have acquired metal-poor gas from outside due to some external processes and are forming a new population of stars.

\subsection{Kinematics} \label{sec: kinematics}
\begin{figure*}
    \centering
    \includegraphics[width = 0.75\textwidth]{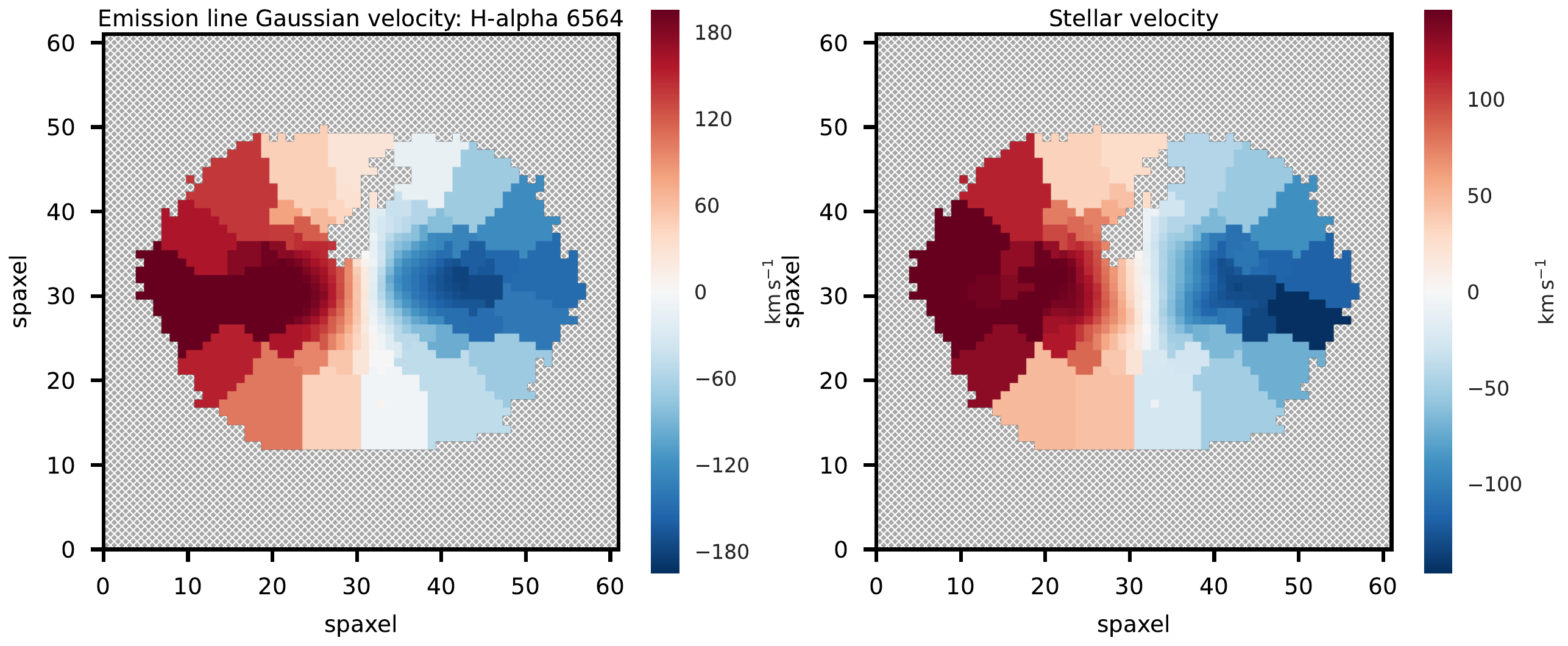}
    \includegraphics[width = 0.75\textwidth]{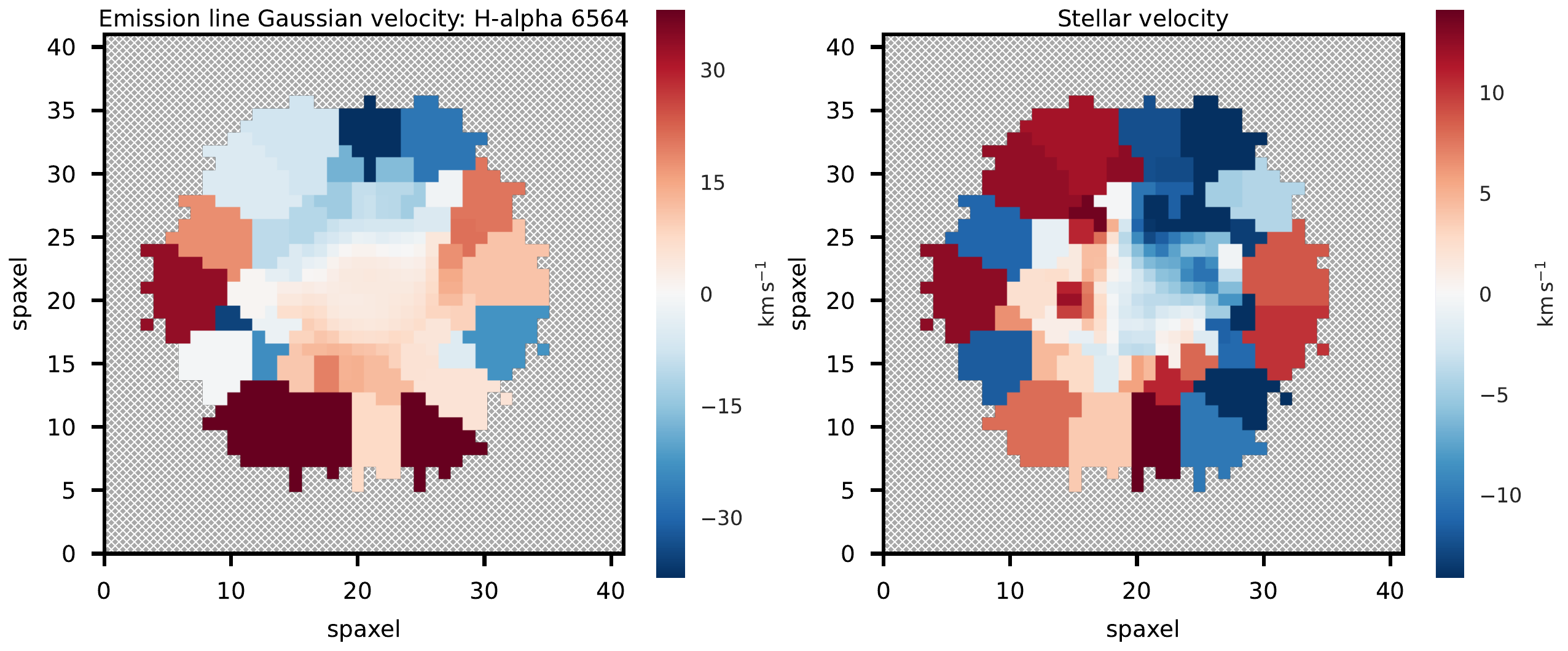}
    \includegraphics[width = 0.75\textwidth]{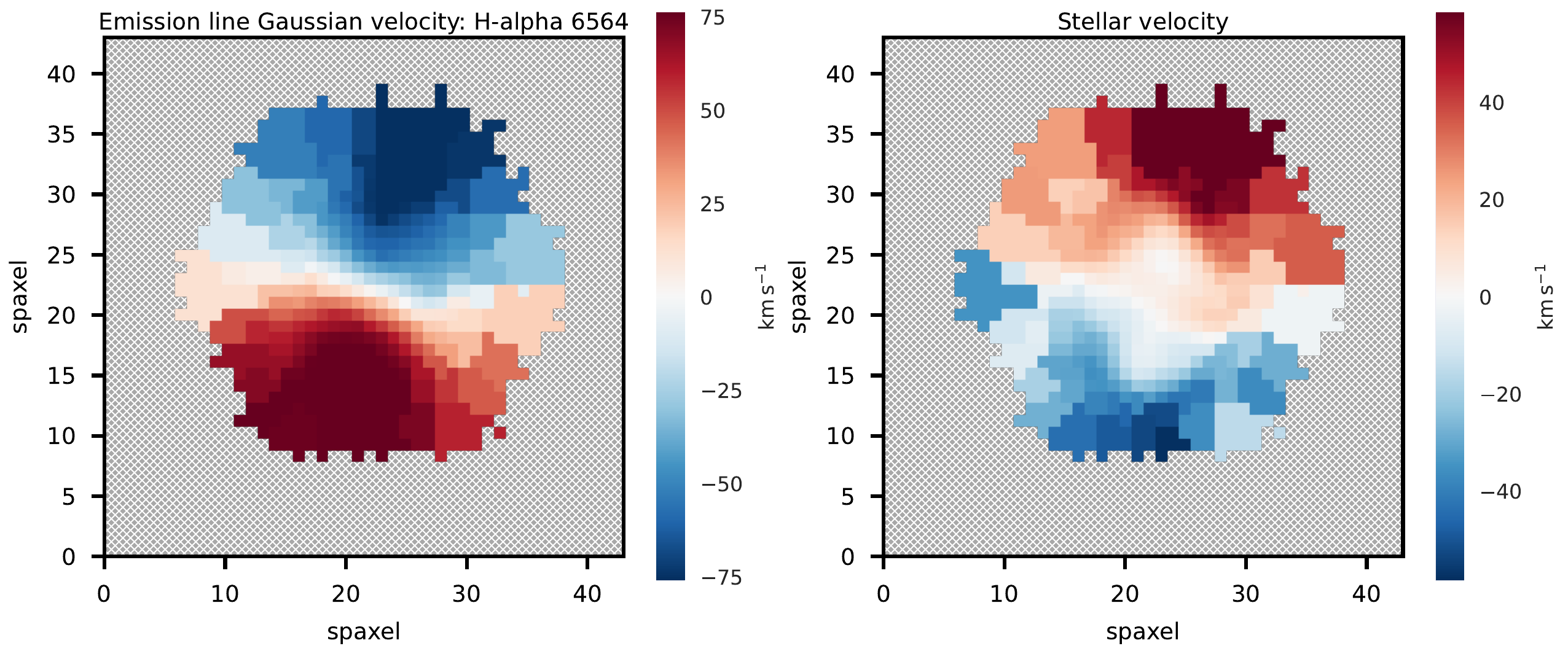}
    \includegraphics[width = 0.75\textwidth]{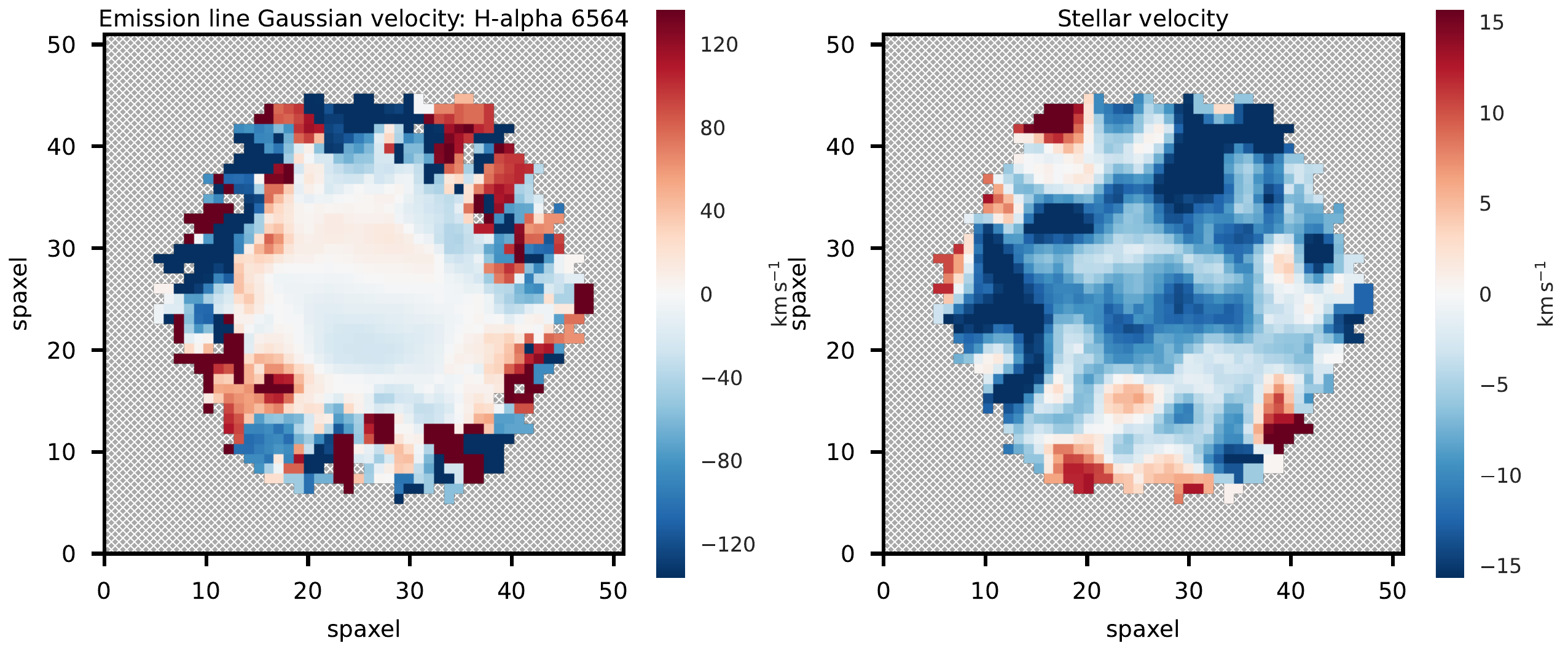}
    \caption{Example gas (left) and stellar (right) velocity maps of regular, complex, counter-rotating, and no rotation for the four categories (from top to bottom). See the text for a description of each category.}.
    \label{fig: velocity_maps}
\end{figure*}
Next, we inspected the kinematic maps of the SF-Es and checked for any disturbances seen in the H$\alpha$ gas- and stellar- velocity maps. In Appendix \ref{sec: imgs}, Figure \ref{fig:fig_A2} shows the gas- and stellar- velocity maps for all the SF-Es. Any disturbance in these maps could signify interaction activity, such as a merger or gas accretion. We have visually examined all the H$\alpha$ gas and stellar velocity maps of SF-Es. Based on this visual inspection, we have divided our sample of SF-Es into four categories:
\begin{itemize}
    \item \textbf{Regular:} These are the galaxies where we can see clear rotational structures in both maps, and the rotational axes are aligned. We have 19 galaxies like this.

    \item \textbf{Complex:} Here, either or both maps show disturbance (do not have any clean rotational structure as we see for the regular category) in the system. Some of these objects have gas and stellar rotational axes misaligned. There are 34 galaxies in this class.
    
    \item \textbf{Counter-rotating:} These objects have a clean rotational structure, but the rotation axes of stars and gas are opposite to each other. There are five objects in this category.
    
    \item \textbf{No rotation:} Only one object, a face-on galaxy, where we can not see evidence of rotation in the stellar and gas maps.
\end{itemize}
In Fig. \ref{fig: velocity_maps}, we have presented one galaxy from each category (regular, complex, counter-rotating, and no rotation from top to bottom). Here, we want to point out that this classification is based only on visual inspection. Sometimes, due to the small inclination angle of the galaxies, we cannot see the velocity structures correctly. For the SF-Es, we have only three objects with inclination angles less than 15\textdegree{}. So, only these three galaxies can be called face-on galaxies. Among them,  the first one is in the no-rotation class (inclination angle 11.\textdegree{}38), the second one is in the regular class (inclination angle 13.\textdegree{}70), and the third one is in the complex class (inclination angle 14.\textdegree{}93). Hence, we believe that the disturbance seen in the velocity maps for the complex category of SF-Es is not caused by the low inclination angle of these galaxies.
\begin{figure}
    \centering
    \includegraphics[width = 0.5\textwidth]{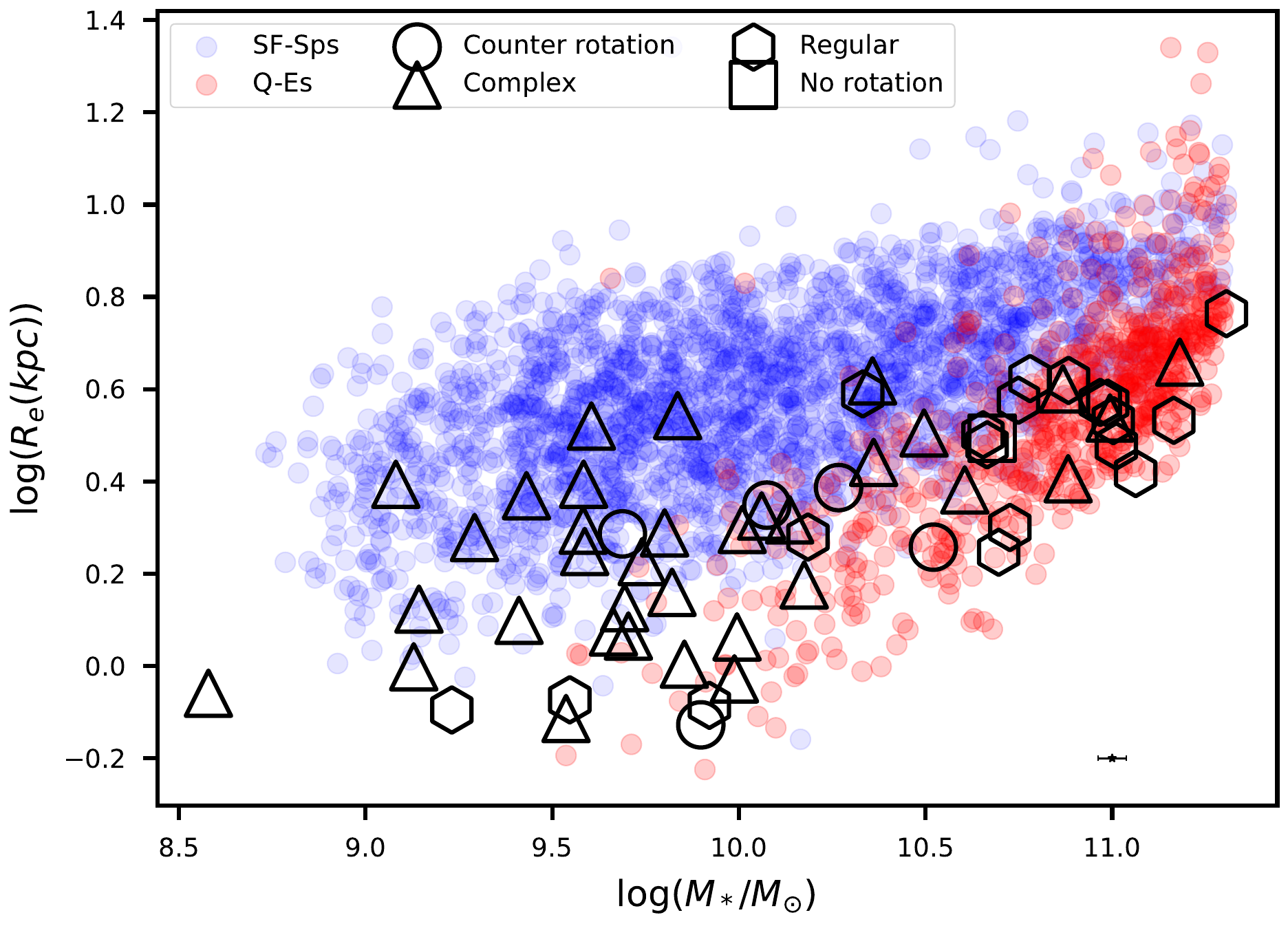}
    \caption{Kinematic disturbance trend of the SF-Es on the size-mass plane. SF-Sps are shown in blue circles, and in red circles, Q-Es are shown. SF-Es are marked with different shapes according to their kinematic disturbance categories (see section \ref{sec: kinematics}). High-mass SF-Es tend to be more regular and most of the low-mass SF-Es are kinematically unsettled. }. 
    \label{fig: size_mass}
\end{figure}

To determine any dependence of the kinematics on the size and mass of the galaxy, we have plotted these objects on the size-mass plane and marked them according to their categories. We can see from Fig. \ref{fig: size_mass} that high-mass SF-Es tend to be more regular than low-mass SF-Es, which tend to be kinematically complex. The reason for this could be the high rotational inertia of massive galaxies, 
 making it difficult to perturb them kinematically.

Further, using a similar argument as \citet{Rathore22}, we have divided our sample of SF-Es as kinematically settled and unsettled galaxies, where kinematically settled galaxies are the regular ones ($19$ galaxies), and kinematically unsettled include complex and counter-rotating galaxies ($34+5=41$ galaxies). So, of the total sample of SF-Es, approximately two-thirds are kinematically unsettled. Recent interaction and/or mergers are likely to be the reasons behind this \citep{Bendo00, Matteo07, Bassett17}, which might also have supplied the fresh gas needed to form new stars. 

\begin{figure}
    \centering
    \includegraphics[width = 0.5\textwidth]{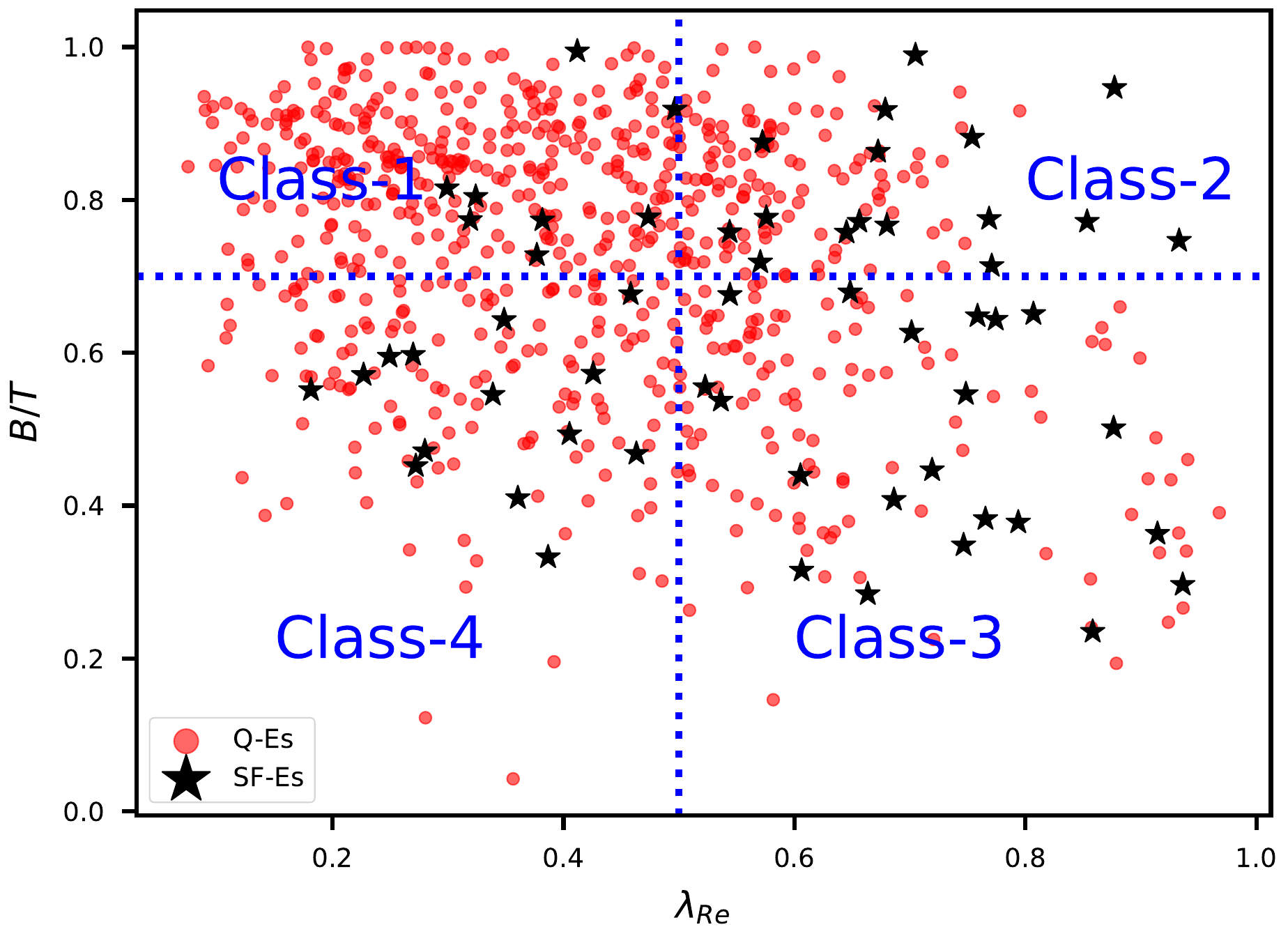}
    \caption{Distribution of SF-Es (as black stars) and Q-Es (as red circles) in the $B/T$ and $\lambda_{Re}$ plane. According to their distribution, we subdivided SF-Es into four classes. Two division criteria ($B/T = 0.7$ and $\lambda_{Re} = 0.5$) are shown by the blue dotted lines.}.
    \label{fig: classification}
\end{figure}

\subsection{Distribution of SF-Es in $B/T-\lambda_{Re}$ plane} \label{sec: age metallicity}
To study the structure and kinematics of the SF-Es, we plot them on the $B/T-\lambda_{Re}$ plane (Fig. \ref{fig: classification}), where we have taken $B/T$ as the proxy for structure and $\lambda_{Re}$ as the proxy for kinematics. Then, we divided them into four classes based on their $B/T$ and $\lambda_{Re}$ as follows:
\begin{itemize}
    \item Class-1 (C1): These objects have $B/T>0.7$ and $\lambda_{Re}<0.5$. These are bulge-dominated, slow rotators; we have eight of them in our primary sample of SF-Es.
    
    \item Class-2 (C2): This class of objects have $B/T>0.7$ and $\lambda_{Re}>0.5$. In the SF-E sample, we have 16 galaxies in this bulge-dominated, fast rotators category.
    
    \item Class-3 (C3): These galaxies have $B/T<0.7$, so they are not totally bulge dominated and $\lambda_{Re}>0.5$, i.e. fast rotators. We have 21 of these SF-Es.
    
    \item Class-4 (C4): Like Class-3, these objects are also not entirely bulge dominated ($B/T<0.7$), but unlike Class-3, they are slow rotators ($\lambda_{Re}<0.5$). There are 14 galaxies in this class of SF-Es.
\end{itemize}
Next, we calculated the weighted mean of the stellar mass and environmental density of these classes, and the following are in decreasing order of stellar mass and environmental density.

Stellar mass (weighted mean): C1 $>$ C2 $>$ C4 $>$ C3

Environmental density (weighted mean): C1 $>$ C3 $>$ C4 $>$ C2
\\
So, the stellar mass of the C1 and C2 is the highest, which is expected as they are bulge-dominated ellipticals. The environmental density of C1 is the highest among all the classes, which is also as expected, as they are bulge dominated,  most massive slow rotators. However, the environmental density of C2 is the lowest, which is entirely unexpected. We will discuss these classes individually in Section~\ref{sec: discussion}.

\begin{figure*}
    \centering
    \includegraphics[width = 0.47\textwidth]{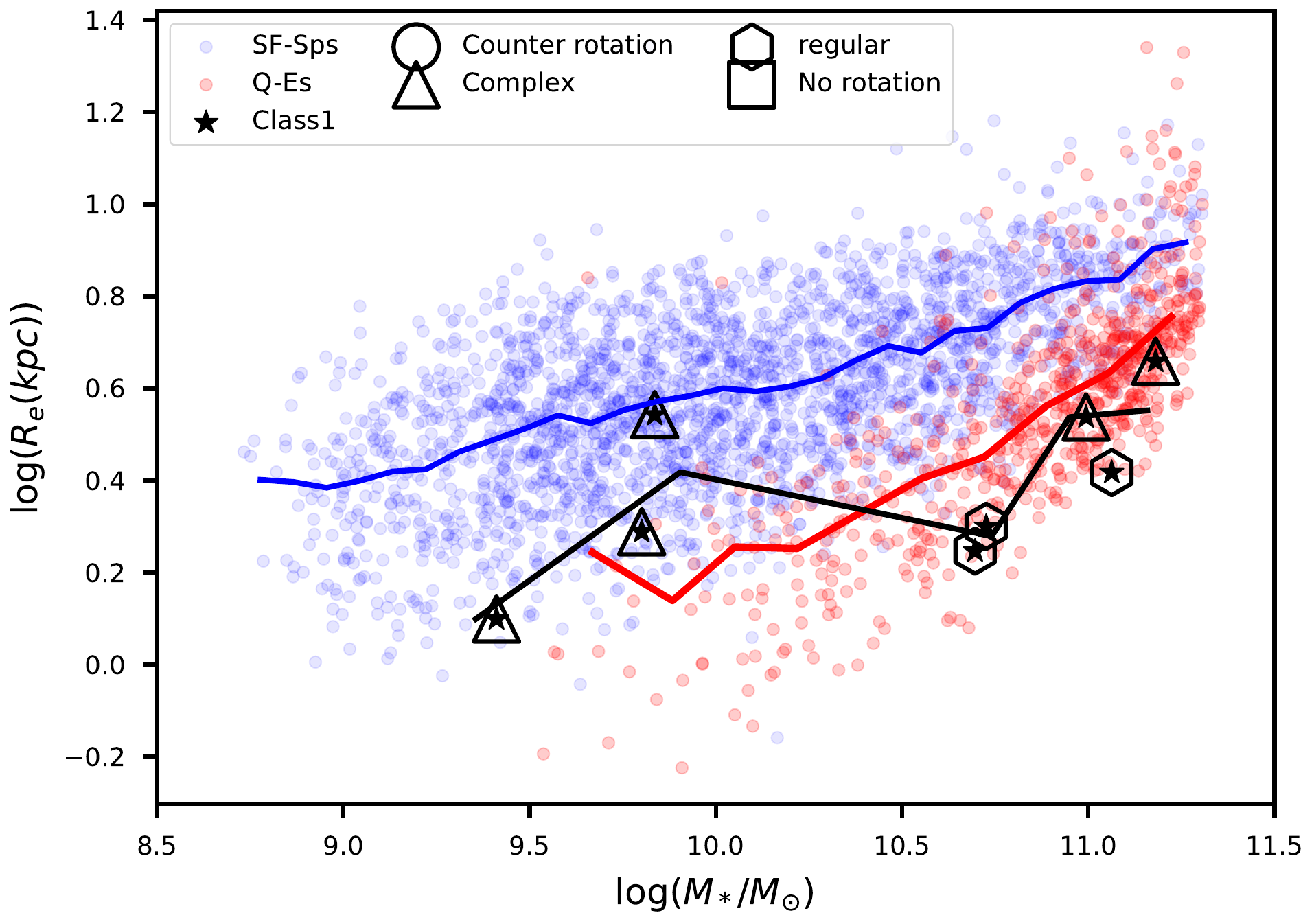}
    \includegraphics[width = 0.47\textwidth]{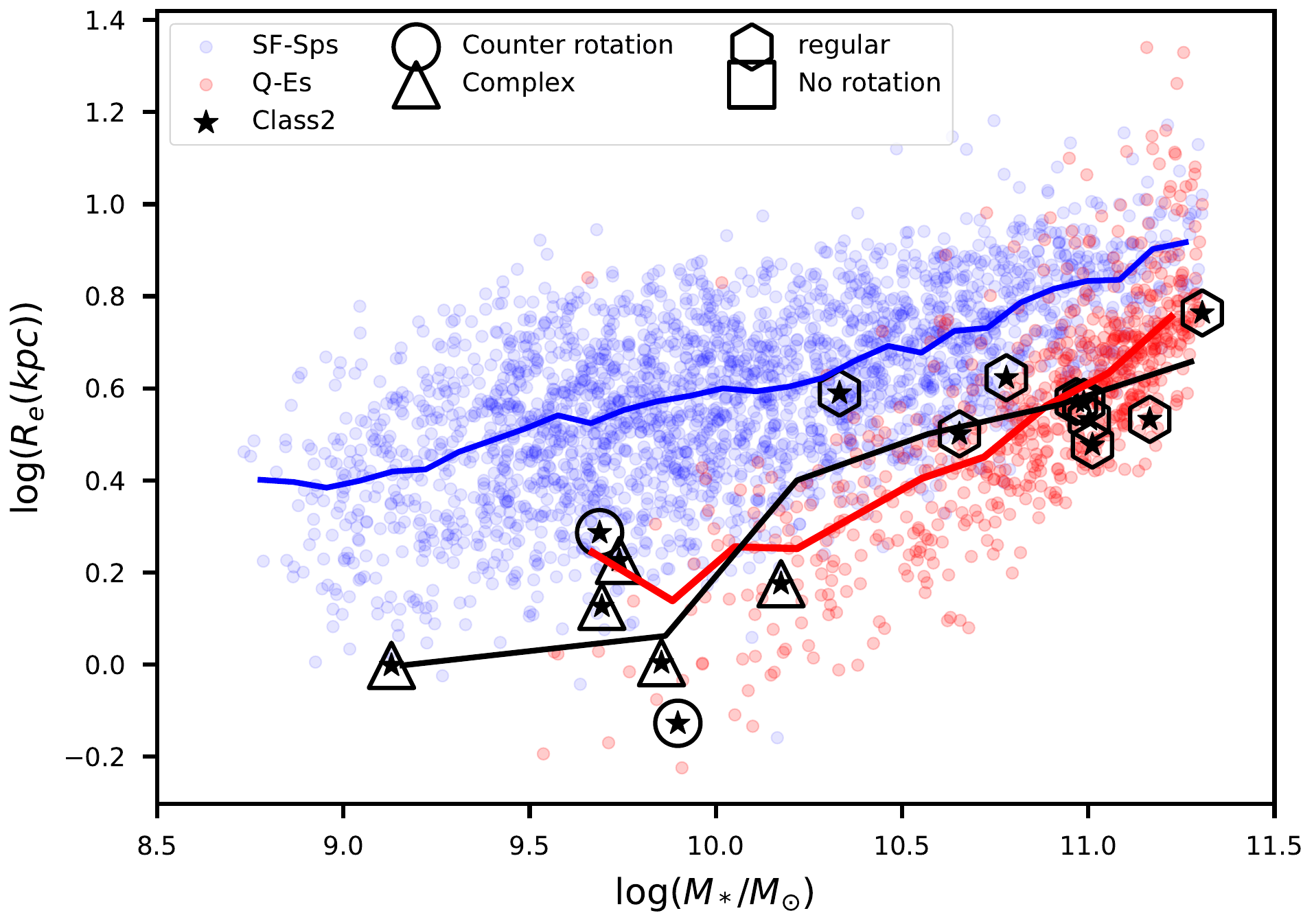}
    \includegraphics[width = 0.47\textwidth]{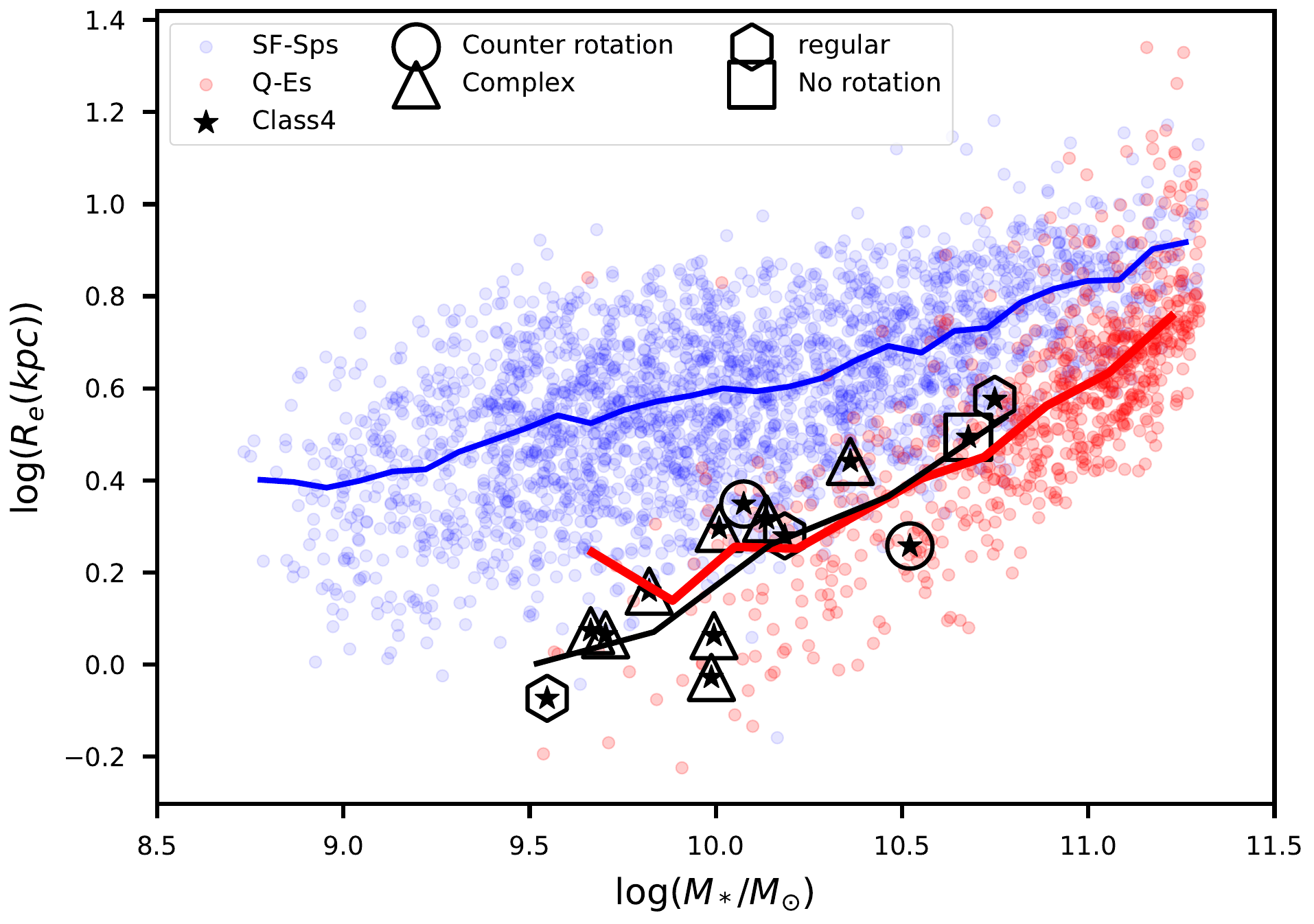}
    \includegraphics[width = 0.47\textwidth]{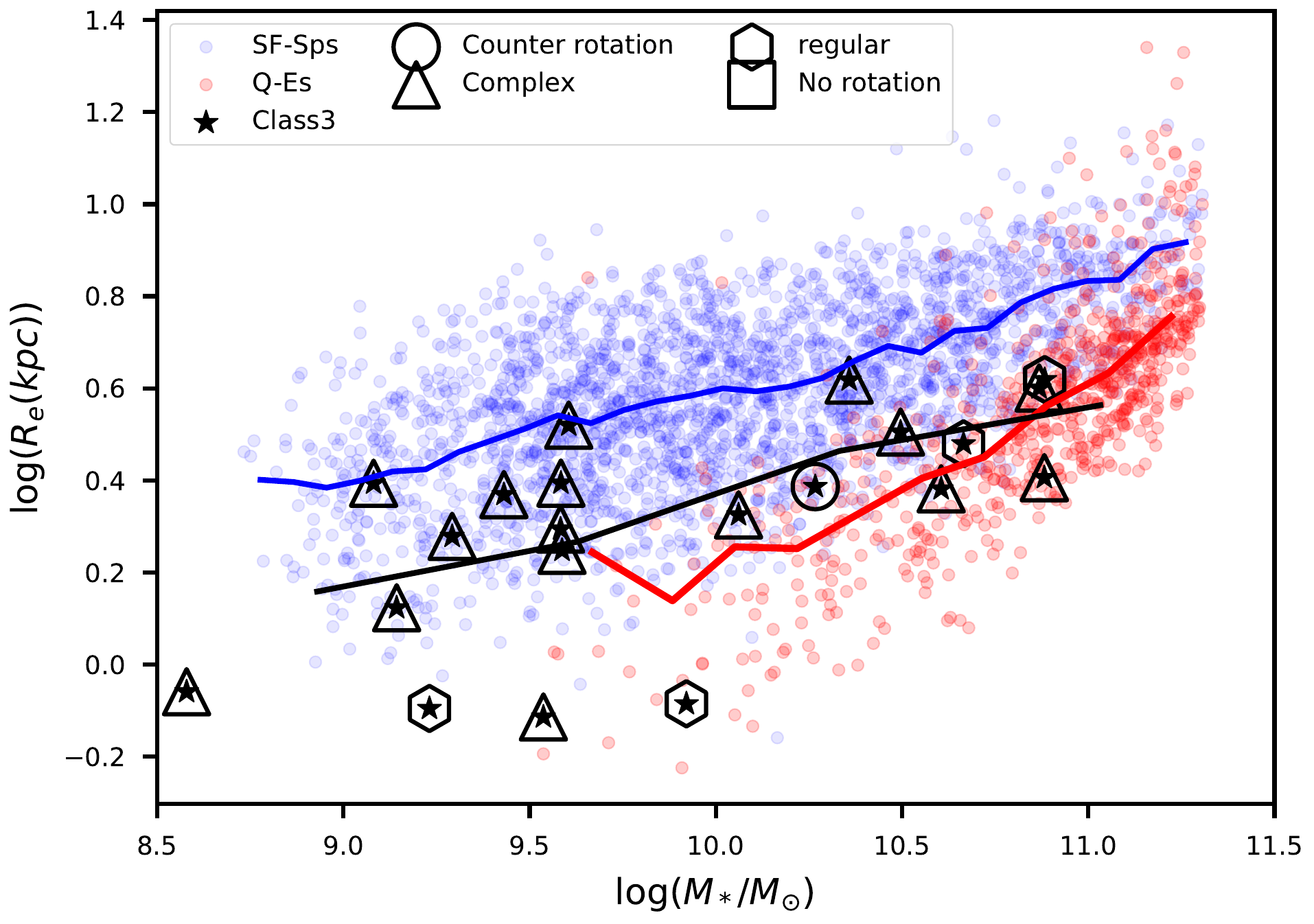}
    \caption{The distribution of the four classes (class-1: upper left, class-2: upper right, class-3: lower right, class-4: lower left) of SF-Es are shown in the size-mass plane. In the background, we have also plotted the control samples of SF-Sps (blue circles) and Q-Es (red circles). The primary sample of SF-Es is shown with black stars. Their kinematic classes are marked with different shapes. The weighted mean trends of the three samples are shown with solid lines.}.
    \label{fig:size_mass_all4}
\end{figure*}

\section{Discussion} \label{sec: discussion}
The Hubble sequence \citep{Hubble1926b} is the morphological classification of galaxies based on visual inspection. It divides galaxies roughly into two broad classes: ETGs and LTGs. ETGs include elliptical and lenticular galaxies where Es are dominated by the bulge and supported by high stellar velocity dispersion, and S0s have fast-rotating extended disks. These ETGs are generally red in color, gas poor with low or very weak star formation, and are found preferentially in high-density environments. On the contrary, LTGs are spiral galaxies dominated by spiral arms in their extended disk. They are generally blue, gas rich, star forming, and found in low-density regions. The Hubble sequence does have its shortcomings. For instance, it considers a simplistic view of the sequential evolution of galaxy morphology from `early type' to `late type'. Also, we cannot correctly classify galaxies that are in transition between ETG and LTG.  Despite these shortcomings, the Hubble sequence is useful for understanding several global galaxy properties and remains widely used in the literature.  In this paper, we analyze a class of outliers: elliptical galaxies that form stars at a rate similar to spiral galaxies present in the SFMS despite being morphologically elliptical. These galaxies are extremely uncommon, but their existence is causing us to question our current understanding of how galaxies evolve. In this paper, we aim to answer some basic questions about this population of star-forming ETGs by studying star-forming ellipticals. The questions we are interested in are:
\begin{itemize}
    \item What are the global properties of these galaxies relative to the two control samples?
    \item What drives star formation in these galaxies?
    \item Where does the gas come from?
    \item Do some external processes trigger star formation?
    \item What are the different evolutionary modes for forming stars in these galaxies?
    \item Can we observationally distinguish between these modes?
\end{itemize}

In our study, only 5\% of ellipticals were SF-Es, consistent with previous studies \citep{Schawinski09, Jeong21} reporting $\sim2-7\%$ of the nearby ETGs are star forming. H$\alpha$ gas and stellar velocity maps of the $\sim67\%$ SF-Es are kinematically unsettled. This includes galaxies with misaligned velocity axes, counter-rotating, and kinematically disturbed velocity maps. Previous surveys like SAURON, SAMI, CALIFA, and ATLAS$^{\rm{3D}}$ have also found that many early-type galaxies (ETGs) show misaligned gas velocity maps with respect to the stellar velocity map \citep{Emsellem04, Sarzi06, Davis11, Barrera15, Bryant19}. The study by \citet{Davis11} revealed that approximately 36\% of fast-rotating ETGs in the ATLAS$^{\rm{3D}}$ have kinematic misalignment. However, this percentage could be higher for star-forming ETGs, as observed in this study. The disturbance in the velocity maps of galaxies indicates external activity like a merger or gas accretion, which can potentially trigger star formation by bringing gas into the system.

Studies of star-forming ETGs have suggested that recent or ongoing interaction with gas-rich neighboring galaxies is one of the main reasons behind the star formation in passively evolving ETGs. Suppose this process is the primary mechanism behind the formation of star-forming ETGs. In that case, we should expect these galaxies to have higher gas content, lower metallicity, and lower stellar population age than the quiescent ETGs. By studying the H$\alpha$ emission line flux, we observe that SF-Es have ionized gas content similar to SF-Sps. \citet{Wei09}, using Green Bank Telescope HI data for 27 E/S0s, found that blue sequence E/S0s have atomic gas masses in the range $10^7-10^{10} M_{\odot}$, which is substantial and comparable to the atomic gas content of spiral and irregular galaxies with similar stellar mass ranges. Their analysis also suggests that many of these galaxies are capable of significant stellar disk growth.

The mass-weighted stellar population age and metallicity of SF-Es are lower than those of Q-Es, as shown in Fig. \ref{fig: age_metallicity}. The D4000 spectral index of SF-Es ranges from 1.2 to 1.8 (see Fig. \ref{fig: D4000}), implying that these galaxies have a young stellar population mixed with an older stellar population or a stellar population with intermediate age. \citet{Jeong21}, inspecting the H$\beta$, Mg b, Fe 5270, and Fe 5335 line strengths, found that star-forming ETGs are more metal poor than quiescent galaxies.

We divided our primary sample of SF-Es into four classes depending on their position on the $B/T-\lambda_{Re}$ plane: Class-1, -2, -3, and -4. Here, we will discuss these four classes individually and try to understand their evolutionary history to know the reason for the high SFR in these galaxies. For this, we have plotted them on the size-mass plane and shown their kinematic nature. In the background of the plots, we have also shown our two control samples of SF-Sps and Q-Es. The weighted mean trend of each sample is shown with a solid line (Fig.~\ref{fig:size_mass_all4}).

The Class-1 objects have $B/T>0.7$ (bulge dominated) and $\lambda_{Re}<0.5$ (slow rotator). Their stellar masses roughly range from $10^9$ to $10^{11}$ M$_\odot$, i.e. they have high masses. The weighted mean of the environmental density is also the highest within these four classes.  So, all the properties of this class are like other high-mass ellipticals except for their high SFR. Figure~\ref{fig:size_mass_all4} (upper left) shows that, out of eight galaxies in this class, three are kinematically regular, and five are disturbed. So, these rare ellipticals could be regular quenched ellipticals in the past. But recently, their star formation has been rejuvenated, likely due to external processes like mergers or accretion.

Class-2 objects are also bulge dominated $(B/T>0.7)$ but fast rotators ($\lambda_{Re}>0.5$). Their mass ranges from $10^9$ to $10^{11}$ M$_\odot$. There are two groups: the high-mass galaxies, which are all kinematically regular, and the relatively lower mass galaxies, which are all kinematically unsettled (see Fig.~\ref{fig:size_mass_all4} upper right). Interestingly, this class of objects has the least weighted mean environmental density. These galaxies could be morphologically transformed spirals. When there is a gas-rich merger, the gas dissipates energy, falls toward the galaxy's center, and creates a compact starburst region, forming new stars \citep{Mihos94}. It has been seen for many fast rotators that their centers are denser than their edges, and they exhibit a steep inner profile \citep{Kormendy09}. This creates a bulge, and the galaxy slowly changes its morphology from LTG to ETG.

According to \citet{Cappellari13a}, there are two separate formation scenarios of the ETGs. The first one is the in situ star formation, where the galaxy grows via minor gas-rich mergers or gas accretion. Due to the potential well of the galaxy, the gas sinks toward the center and forms a bulge, which is responsible for the star-formation quenching of the galaxy and disk fading \citep{Cheung12, Fang13}. Because of this process, an LTG morphologically changes to an ETG. The second process predominantly involves dry mergers, leaving the stellar population unchanged \citep{Bezanson09, Naab09, Hopkins10}. These two processes are dependent on the environmental density of the galaxies. The first process is dominant in low-density regions, and the second dominates in high-density regions. Also, the ETGs formed via the first process tend to be fast rotators. However, the second method creates slow rotator ETGs. Multiple studies have found that because of the environmental effect in clusters, spirals become passive and transform into S0s \citep{Smith05, Boselli06, Moran07}.

Hence, there is a two-channel build-up process for ETGs. One channel is the bulge growth and quenching route, and here, spiral galaxies are gradually replaced by early-type fast rotators. This channel could form our Class-2 galaxies, where their star formation is still active, and they are in the process of quenching. The second route is the dry merger-growth route, and the mass increase in the slow rotators illustrates it. Our Class-1 objects could result from this second channel, but clearly, their merger growth was not completely dry and star formation was at least partially rejuvenated during the merger process.

Class-3 objects have $B/T<0.7$, i.e. they are not entirely bulge dominated and have $\lambda_{Re}>0.5$, so they are fast rotators. Their weighted mean environmental density is also higher next to the class-1 objects. Most of these objects are kinematically unsettled, and their mean trend (see Fig.~\ref{fig:size_mass_all4} lower left) is also between SF-Sps and Q-Es. Our interpretation for this class is that these were quenched S0s in the past, and they have rejuvenated their star formation in recent evolutionary history. Almost all the objects here are kinematically complex, indicating recent interactions like merger and/or gas accretion, which increased their gas content fueling the star formation. Some of the galaxies in this class could be morphologically transformed spirals, which we have discussed above.

According to the mean trend, class-4 is similar to the ellipticals (see Fig.\ref{fig:size_mass_all4} lower right panel). But their $B/T<0.7$, so these objects have a disk and are not entirely bulge dominated. Almost all of them are kinematically unsettled, and their stellar mass ranges from $10^{9.5}$ to $10^{11}$ M$_\odot$. These could be the low-redshift counterparts of the red nuggets \citep{Damjanov09}, which have acquired, over time, a star-forming disk via the ‘Disk-cloaking’ process.

Red nuggets are a class of massive, ultra-compact, quiescent galaxies found in $z\sim1-2$ but nonexistent in the local universe, first reported by \citet{Daddi05}. These galaxies are small in size, but their stellar mass is similar to the giant ellipticals \citep{Longhetti07, Toft07, Cimatti08, Damjanov09}. Studies \citep{Trujillo09, Saulder15} based on SDSS show that red nuggets are absent from the local universe. On the other hand, since $z\sim1.5$, the fraction of elliptical galaxies has increased by four times in the last 9 Gyrs \citep[$z\sim1.5$;][]{Buitrago13}. The similarity of mass and morphology could connect red nuggets to elliptical galaxies. There is a process called the E-to-E process, in which red nuggets go through a size growth and morphologically change to elliptical galaxies in the present-day universe. In addition, there is a finite possibility that red nuggets have not evolved into quiescent galaxies. They can restart their star formation by accreting gas and stellar clusters around them. It would be a mistake only to assume that red nuggets evolve into ellipticals at present as they are similar in morphology. They can also transform into nonspheroidal systems like S0s and spirals. \citet{Graham2013} suggests an alternative evolutionary mechanism known as ‘Disk-cloaking’ where red nuggets go through minor mergers and accretion \citep{Navarro1991, Birnboim03, Pichon11} and experience significant disk-growth, and they are now present in the embedded bulges in S0 and some early-type spiral galaxies. It is possible that this process is the main mechanism responsible for the formation of Class-4 galaxies.

The above results motivate the need to revise the classic Hubble tuning fork diagram \citep{Hubble36}.  A proposed revision of the same, known as the ATLAS-3D comb morphology diagram, was suggested by \citet{Cappellari11b}. The shape of this new diagram is not a symmetric fork or a trident but an asymmetric comb. The ETGS are replaced parallel to the LTGs from the handle of the tuning fork. The main features of this revised morphology diagram are as follows:
\begin{itemize}
    \item A clear distinction between the slow and fast rotators and the link between the S0s and the fast rotators. This takes into account galaxy kinematics, which plays a vital role in galaxy evolution \citep{Cappellari16}. Different IFS and redshift evolution studies independently and consistently show that the evolutionary paths of fast and slow rotators are distinct, one being driven by gas accretion and the other by dry mergers.
    
    \item The spirals and S0s are parallel to each other in this revised classification, indicating the close link between spirals and fast rotator ETGs \citep{van76}.
    
    \item It incorporates the proposal by \citet{Kormendy96} that there should be a link between the S0s and the disky ellipticals.
\end{itemize}
The last two points imply that galaxies in their lifetime change their morphology. All of our results with the comb morphology diagram, and they both emphasize that many of the essential properties of galaxies, like the gas content and stellar population age, vary along the sequence from ETGs to LTGs and are also shared by the fast and slow rotators populations of ETGs.

\section{Summary and conclusions} \label{sec: summary}

In this work, we have identified and studied a rare class of star-forming ellipticals (SF-Es). Combining data from the DL17 and the Salim catalogs, we identified 59 star-forming elliptical galaxies in the SDSS-MaNGA. Being an IFS survey, MaNGA can provide data on local scales across the 2D face of a galaxy. Using MaNGA Pipe3D and other catalogs, we studied this enigmatic class of SF-Es, both on resolved and global scales. Traditionally, elliptical galaxies are thought to be non or very weakly star forming. So, a detailed study of the characteristics of these galaxies can give us information about galaxy evolution and how galaxies go through different morphological states and move across the star-forming main sequence.

We study the distribution of the galaxies based on several parameters like stellar mass, SFR, effective radius, $B/T$, $\lambda_{Re}$, stellar age, and metallicity across the three samples on the global scale. We also checked the resolved maps of gas velocity, stellar velocity, H$_\alpha$ emission, and D4000 for the 59 SF-Es to get information on the local scale. The following are our main findings:
\begin{itemize}
    \item The stellar mass distribution shows that SF-Es, on average, have lower stellar mass than Q-Es. The mass distribution of the SF-Es is similar to the SF-Sps. The number of star-forming galaxies amongst spirals and ellipticals declines roughly after $10^{10}$ M$_{\odot}$. Also, SF-Sps and SF-Es are distributed in similar environment-density regions. On the other hand, Q-Es reside in higher density environments. This is consistent with the idea that mass and environmental quenching play an important role in galaxy evolution.
    
    \item $\lambda_{Re}$ for most of the SF-Sps are greater than 0.6, i.e., fast rotators. The number of Q-Es decreases rapidly after 0.6, i.e. Q-Es are mostly slow rotators. Our primary sample of SF-Es contains slow as well as fast rotators. This suggests that using galaxy kinematics is essential for a complete understanding of galaxy evolution processes.
    
    \item We considered H$\alpha$ emission line flux as a measure of the gas content (ionized) of the galaxy. We find that the gas content of SF-Sps is high, and Q-Es have very low gas content. Apart from some galaxies, most SF-Es have H$\alpha$ emissions similar to SF-Es. H$\alpha$ maps of SF-Es show that the bulk of the H$\alpha$ emission comes from the central region of the galaxies.
    
    \item The D4000 spectral index can separate old (D4000~1.6-2.0) and young (D4000~1.1-1.4) stellar populations in the galaxies. The distribution of spectral index D4000 shows that SF-Sps have the young stellar population, Q-Es have the old stellar population, and the primary sample of SF-Es has a mixture of old and young stellar populations. The D4000 maps of SF-Es show three types of galaxies. The maximum number of galaxies show a young stellar population at the center and an old population in the outer regions; some show an old population at the center and a young population outside, and very few show a young population of stars distributed all over the galaxy.
    
    \item When two samples of ellipticals, star-forming and quenched, were plotted in the mass-weighted stellar population age-metallicity plane, it shows that the SF-Es have systematically lower metallicity and age of the stellar population compared to the Q-Es. So SF-Es seem to have acquired metal-poor gas in the recent past.
    
    \item The complex velocity structure of the velocity maps indicates disturbances in the system, possibly because of some external or internal interactions in the galaxies. Two-thirds ($\sim$ 67\%) of the SF-Es are kinematically unsettled, implying recent interactions and mergers. Minor mergers could be a potent driver of rejuvenation in elliptical galaxies by bringing metal-poor gas into the galaxy.

    \item We then subdivided our sample of SF-Es into four classes based on their $B/T$ and $\lambda_{Re}$ values, and every class of galaxies has its own evolutionary history and mode of formation. Class-1 objects are bulge dominated (B/T $>0.7$), slow rotators $(\lambda_{Re}<0.5)$. These galaxies are classical rejuvenated ellipticals. Class-2 objects are also bulge dominated ($B/T >0.7$) but fast rotators $(\lambda_{Re}>0.5)$. These could be morphologically transformed spiral galaxies. Class-3 galaxies are not fully bulge dominated ($B/T <0.7$), fast rotators $(\lambda_{Re}>0.5)$. These could be rejuvenated S0s, fading spirals, or a mixture of both. Class-4 galaxies are also not fully bulge-dominated ($B/T <0.7$) but slow rotators $(\lambda_{Re}<0.5)$. These objects could be similar to red nuggets, which have acquired a star-forming disk.
\end{itemize}

All the above findings imply that these intriguing and rare galaxies do not belong to a homogeneous class of objects. There could be different underlying processes responsible for the formation of SF-Es. Also, our results support the comb morphology diagram, which is a revision of the classical Hubble tuning fork diagram. The comb morphology diagram emphasizes three points: parallelism between spirals and S0s, the link between S0s and ellipticals and the distinction between fast and slow rotators.

Further in-depth studies for larger samples of galaxies, preferably volume-limited, spanning across stellar mass, morphology, and redshift, will be needed to draw a complete picture of this elusive class of star-forming ellipticals. HI and molecular gas observations will provide complementary data in these types of studies. Such observations can provide accurate estimates of the gas content and insights into the efficiency and time scale of star formation in these systems.

\section*{acknowledgments}

We thank the anonymous referee whose comments and suggestions have improved both the content and presentation of this paper.
P.B. and Y.W. acknowledge the support of the Department of Atomic Energy, Government of India, under project no. 12-R\&D-TFR5.02-0700. This research made use of Astropy,\footnote{\url{http://www.astropy.org}} a community-developed core Python package for Astronomy \citep{astropy:2013, astropy:2018, Astropy22}.

Funding for the Sloan Digital Sky Survey IV has been provided by the Alfred P. Sloan Foundation, the U.S. Department of Energy Office of Science, and the Participating Institutions. SDSS-IV acknowledges support and resources from the Center for High-Performance Computing at the University of Utah. The SDSS website is www.sdss.org. SDSS-IV is managed by the Astrophysical Research Consortium for the Participating Institutions of the SDSS Collaboration, including the Brazilian Participation Group, the Carnegie Institution for Science, Carnegie Mellon University, Center for Astrophysics-Harvard \& Smithsonian, the Chilean Participation Group, the French Participation Group, Instituto de Astrof\'isica de Canarias, The Johns Hopkins University, Kavli Institute for the Physics and Mathematics of the Universe (IPMU) / University of Tokyo, the Korean Participation Group, Lawrence Berkeley National Laboratory, Leibniz Institut f\"ur Astrophysik Potsdam (AIP),  Max-Planck-Institut f\"ur Astronomie (MPIA Heidelberg), Max-Planck-Institut f\"ur Astrophysik (MPA Garching), Max-Planck-Institut f\"ur Extraterrestrische Physik (MPE), National Astronomical Observatories of China, New Mexico State University, New York University, University of Notre Dame, Observat\'ario Nacional / MCTI, The Ohio State University, Pennsylvania State University, Shanghai Astronomical Observatory, United Kingdom Participation Group, Universidad Nacional Aut\'onoma de M\'exico, University of Arizona, University of Colorado Boulder,The  University of Oxford, University of Portsmouth, University of Utah, University of Virginia, University of Washington, University of Wisconsin, Vanderbilt University, and Yale University.

The Legacy Surveys consist of three individual and complementary projects: the Dark Energy Camera Legacy Survey (DECaLS; Proposal ID \#2014B-0404; PIs: David Schlegel and Arjun Dey), the Beijing-Arizona Sky Survey (BASS; NOAO Prop. ID \#2015A-0801; PIs: Zhou Xu and Xiaohui Fan), and the Mayall z-band Legacy Survey (MzLS; Prop. ID \#2016A-0453; PI: Arjun Dey). DECaLS, BASS and MzLS together include data obtained, respectively, at the Blanco telescope, Cerro Tololo Inter-American Observatory, NSF’s NOIRLab; the Bok telescope, Steward Observatory, University of Arizona; and the Mayall telescope, Kitt Peak National Observatory, NOIRLab. The Legacy Surveys project is honored to be permitted to conduct astronomical research on Iolkam Du’ag (Kitt Peak), a mountain with particular significance to the Tohono O’odham Nation.

NOIRLab is operated by the Association of Universities for Research in Astronomy (AURA) under a cooperative agreement with the National Science Foundation.

This project used data obtained with the Dark Energy Camera (DECam), which was constructed by the Dark Energy Survey (DES) collaboration. Funding for the DES Projects has been provided by the U.S. Department of Energy, the U.S. National Science Foundation, the Ministry of Science and Education of Spain, the Science and Technology Facilities Council of the United Kingdom, the Higher Education Funding Council for England, the National Center for Supercomputing Applications at the University of Illinois at Urbana-Champaign, the Kavli Institute of Cosmological Physics at the University of Chicago, Center for Cosmology and Astro-Particle Physics at the Ohio State University, the Mitchell Institute for Fundamental Physics and Astronomy at Texas A\&M University, Financiadora de Estudos e Projetos, Fundacao Carlos Chagas Filho de Amparo, Financiadora de Estudos e Projetos, Fundacao Carlos Chagas Filho de Amparo a Pesquisa do Estado do Rio de Janeiro, Conselho Nacional de Desenvolvimento Cientifico e Tecnologico and the Ministerio da Ciencia, Tecnologia e Inovacao, the Deutsche Forschungsgemeinschaft and the Collaborating Institutions in the Dark Energy Survey. The Collaborating Institutions are Argonne National Laboratory, the University of California at Santa Cruz, the University of Cambridge, Centro de Investigaciones Energeticas, Medioambientales y Tecnologicas-Madrid, the University of Chicago, University College London, the DES-Brazil Consortium, the University of Edinburgh, the Eidgenossische Technische Hochschule (ETH) Zurich, Fermi National Accelerator Laboratory, the University of Illinois at Urbana-Champaign, the Institut de Ciencies de l’Espai (IEEC/CSIC), the Institut de Fisica d’Altes Energies, Lawrence Berkeley National Laboratory, the Ludwig Maximilians Universitat Munchen and the associated Excellence Cluster Universe, the University of Michigan, NSF’s NOIRLab, the University of Nottingham, the Ohio State University, the University of Pennsylvania, the University of Portsmouth, SLAC National Accelerator Laboratory, Stanford University, the University of Sussex, and Texas A\&M University.

BASS is a key project of the Telescope Access Program (TAP), which has been funded by the National Astronomical Observatories of China, the Chinese Academy of Sciences (the Strategic Priority Research Program “The Emergence of Cosmological Structures” Grant \# XDB09000000), and the Special Fund for Astronomy from the Ministry of Finance. The BASS is also supported by the External Cooperation Program of the Chinese Academy of Sciences (Grant \# 114A11KYSB20160057), and the Chinese National Natural Science Foundation (Grant \# 11433005).
The Legacy Survey team makes use of data products from the Near-Earth Object Wide-field Infrared Survey Explorer (NEOWISE), which is a project of the Jet Propulsion Laboratory/California Institute of Technology. NEOWISE is funded by the National Aeronautics and Space Administration. The Legacy Surveys imaging of the DESI footprint is supported by the Director, Office of Science, Office of High Energy Physics of the U.S. Department of Energy under Contract No. DE-AC02-05CH1123, by the National Energy Research Scientific Computing Center, a DOE Office of Science User Facility under the same contract, and by the U.S. National Science Foundation, Division of Astronomical Sciences under Contract No. AST-0950945 to NOAO.

\section*{Data availability}
All the data used in this study are publicly available. The catalog of star-forming ellipticals (SF-Es), star-forming spirals (SF-Sps), and quenched ellipticals (Q-Es) that were constructed are available in the online version of this paper as supplementary material.

%

\vspace{5mm}
\facilities{SDSS-MaNGA}


\software{TOPCAT \citep{Taylor05}, astropy \citep{astropy:2013, astropy:2018, Astropy22}}



\appendix \label{appn}

\section{Data Tables} \label{sec: table}
Tables \ref{tab: SFE_Table}, \ref{tab:QE_Table}, and \ref{tab:SFSp_Table} represent the various parameters of the first 10 galaxies from the samples of SF-Es, Q-Es, and SF-Sps, respectively.
\begin{splitdeluxetable*}{ccccccccccBcccccccccc}
\tabletypesize{\scriptsize}
\tablewidth{0pt} 
\tablecaption{Table of the primary sample of SF-Es. \label{tab: SFE_Table}}
\tablehead{
\colhead{PLATEIFU} & \colhead{RA (J2000)}& \colhead{DEC (J2000)} & \colhead{z} & \colhead{TTYPE} & \colhead{P\_LTG} & \colhead{P\_S0} & \colhead{VC} & \colhead{$\log(\frac{M_\ast}{M_\odot})$} &
\colhead{$\log$(SFR)} \\
\colhead{} & \colhead{(degree)} &  \colhead{(degree)} & \colhead{} & \colhead{} & \colhead{} & \colhead{} & \colhead{} & \colhead{} & \colhead{($M_\odot$ yr$^{-1}$)} &
\colhead{$\log(\Sigma)$} & \colhead{$\lambda_{Re}$} & \colhead{H$\alpha$} & \colhead{D4000} & \colhead{log(age)} & \colhead{$\log(z/z_{\odot})$} & \colhead{B/T} & \colhead{Re} & \colhead{PA} & 
\colhead{Kinematic category} \\
 \colhead{} & \colhead{} & \colhead{} & \colhead{} & \colhead{} & \colhead{} & \colhead{} & \colhead{} & \colhead{} & \colhead{} & \colhead{(Mpc$^{-2}$)} & \colhead{} &\colhead{$(10^{-16}erg/s/cm^2)$} & \colhead{} & \colhead{(yr)} & \colhead{} & \colhead{} & \colhead{(kpc)} & \colhead{(degree)} & \colhead{}
} 
\colnumbers
\startdata 
8622-3704 & 351.2176 & 14.1389 & 0.04 & -1.06 & 0.04 & 0.43 & 1 & 10.075 & -0.005 & 0.67 & 0.46 & 0.78 & 1.37 & 9.71 & -0.3 & 0.68 & 2.23 & -87.62 & 2 \\
    12700-3704 & 2.2593 & 15.1266 & 0.04 & -1.47 & 0.02 & 0.44 & 1 & 9.821 & -0.797 & -0.63 & 0.35 & 0.22 & 1.39 & 9.63 & -0.33 & 0.64 & 1.44 & -20.67 & 1 \\
    8089-1901 & 6.8239 & 15.9126 & 0.04 & -1.5 & 0.01 & 0.48 & 1 & 9.547 & -0.312 & -0.37 & 0.25 & 1.83 & 1.17 & 9.61 & -0.6 & 0.6 & 0.84 & 36.14 & 0 \\
    9183-6102 & 122.1352 & 38.9053 & 0.01 & -1.85 & 0.12 & 0.47 & 1 & 8.579 & -1.821 & -0.97 & 0.52 & 0.03 & 1.48 & 9.6 & -0.41 & 0.56 & 0.87 & 84.4 & 1 \\
    8147-1902 & 117.1775 & 26.5398 & 0.02 & -0.22 & 0.04 & 0.38 & 1 & 9.231 & -0.929 & -0.43 & 0.76 & 0.66 & 1.34 & 9.51 & -0.42 & 0.65 & 0.8 & 16.5 & 0 \\
    8948-6103 & 165.1983 & 49.9019 & 0.03 & -2.34 & 0.0 & 0.4 & 1 & 9.988 & -0.658 & 0.32 & 0.27 & 0.27 & 1.65 & 9.81 & -0.14 & 0.6 & 0.94 & 32.27 & 1 \\
    8997-12705 & 172.1561 & 50.8274 & 0.03 & -1.76 & 0.01 & 0.42 & 1 & 10.696 & 0.103 & 0.24 & 0.38 & 0.87 & 1.46 & 9.45 & -0.03 & 0.73 & 1.76 & -13.89 & 0 \\
    9184-1901 & 119.3913 & 33.1215 & 0.02 & -1.28 & 0.01 & 0.38 & 1 & 9.411 & -1.032 & 0.63 & 0.3 & 0.27 & 1.37 & 9.65 & -0.45 & 0.82 & 1.26 & -84.87 & 1 \\
    10214-1902 & 121.9883 & 34.2273 & 0.02 & -3.07 & 0.0 & 0.45 & 1 & 9.898 & -0.651 & -1.35 & 0.57 & 0.72 & 1.53 & 9.57 & -0.1 & 0.72 & 0.75 & -19.25 & 2 \\
    8249-9101 & 136.4765 & 46.2591 & 0.05 & -0.56 & 0.06 & 0.3 & 1 & 10.78 & 0.203 & -0.32 & 0.66 & 0.24 & 1.44 & 9.72 & -0.16 & 0.77 & 4.2 & 41.31 & 0 \\
\enddata
\tablecomments{ 
    \lq PLATEIFU\rq: plate-ifu of the MaNGA datacube,
    \lq RA (J2000)\rq: right ascension,
    \lq DEC (J2000)\rq: declination,
    \lq z\rq: SDSS redshift,
    \lq TTYPE\rq: Hubble TType value,
    \lq P\_LTG\rq: probability of a galaxy being an LTG rather than ETG,
    \lq P\_S0\rq: probability of a galaxy being an S0 as opposed to being elliptical,
    \lq VC\rq: visual classification number (1 for Es, 2 for S0s and 3 for LTGs),
    \lq $\log(\frac{M_\ast}{M_\odot})$\rq: logarithm of the stellar mass of the galaxy relative to the solar mass,
    \lq log(SFR)\rq: logarithm of the star-formation rate of the galaxy,
    \lq $\log(\Sigma)$\rq: logarithm of the environmental density,
    \lq $\lambda_{Re}$\rq: spin parameter,
    \lq H$\alpha$\rq: logarithm of the H$\alpha$ emission line Gaussian flux,
    \lq D4000\rq: D4000 spectral index,
    \lq log(age)\rq: logarithm of the mass-weighted stellar population age,
    \lq $\log(\frac{z}{z_{\odot}})$\rq: logarithm of the mass-weighted stellar population metallicity,
    \lq $B/T$\rq: bulge to total luminosity ratio,
    \lq $R_e$\rq: half-light semi-major axis,
    \lq PA\rq: position angle,
    \lq Kinematic category\rq: kinematic category based on visual inspection of stellar and H$\alpha$ velocity maps ($0$: regular, $1$: complex, $2$: counter-rotating, $3$: no rotation).
    TTYPE, P\_LTG, P\_S0, and VC have been taken from the MaNGA DR17 deep-learning morphology catalog. Stellar mass, SFR and redshift have been taken from the GALEX-SDSS-WISE LEGACY catalog. $B/T$ and PA have been taken from the $r$-band S\'{e}rsic+Exponential fits to the 2D Surface profiles of the MaNGA PyMorph DR17 catalog. Typical errors (median error across all galaxies): $\Delta$ TTYPE $= 0.60$, $\Delta P\_LTG = 0.02$,  $\Delta P\_S0 = 0.16$, $\Delta\log(\frac{M_\ast}{M_\odot}) = 0.04$, $\Delta\log(SFR) = 0.11$, $\Delta \lambda_{Re} = 0.03$, $\Delta H\alpha = 0.15$, $\Delta D4000 = 0.02$, $\Delta \log(age) = 0.03$, $\Delta \log(z/z_{\odot}) = 0.03$. The full table is available online.}
\end{splitdeluxetable*}

\begin{splitdeluxetable*}{ccccccccccBccccccccc}
\tabletypesize{\scriptsize}
\tablewidth{0pt} 
\tablecaption{Table of the control sample of Q-Es. \label{tab:QE_Table}}
\tablehead{
\colhead{PLATEIFU} & \colhead{RA (J2000)}& \colhead{DEC (J2000)} & \colhead{z} & \colhead{TTYPE} & \colhead{P\_LTG} & \colhead{P\_S0} & \colhead{VC} & \colhead{$\log(\frac{M_\ast}{M_\odot})$} &
\colhead{$\log$(SFR)} \\
\colhead{} & \colhead{(degree)} &  \colhead{(degree)} & \colhead{} & \colhead{} & \colhead{} & \colhead{} & \colhead{} & \colhead{} & \colhead{($M_\odot$ yr$^{-1}$)} &
\colhead{$\log(\Sigma)$} & \colhead{$\lambda_{Re}$} & \colhead{H$\alpha$} & \colhead{D4000} & \colhead{log(age)} & \colhead{$\log(z/z_{\odot})$} & \colhead{B/T} & \colhead{Re} & \colhead{PA} \\
 \colhead{} & \colhead{} & \colhead{} & \colhead{} & \colhead{} & \colhead{} & \colhead{} & \colhead{} & \colhead{} & \colhead{} & \colhead{(Mpc$^{-2}$)} & \colhead{} &\colhead{$(10^{-16}erg/s/cm^2)$} & \colhead{} & \colhead{(yr)} & \colhead{} & \colhead{} & \colhead{(kpc)} & \colhead{(degree)}
} 
\colnumbers
\startdata 
12769-9101 & 6.0452 & -1.0203 & 0.06 & -1.6 & 0.02 & 0.14 & 1 & 11.306 & -1.178 & 0.78 & 0.29 & 0.03 & 1.55 & 9.89 & 0.12 & 0.85 & 9.99 & -11.16 \\
    8156-6101 & 54.5971 & -1.1841 & 0.05 & -3.4 & 0.0 & 0.1 & 1 & 11.132 & -1.05 & -0.71 & 0.37 & 0.01 & 1.79 & 9.9 & -0.04 & 0.81 & 5.05 & 24.12 \\
    12685-6104 & 329.7875 & -6.9809 & 0.06 & -2.78 & 0.0 & 0.22 & 1 & 10.931 & -1.205 & -0.14 & 0.25 & 0.01 & 1.77 & 9.89 & -0.14 & 0.81 & 3.77 & 31.72 \\
    12685-3702 & 329.5313 & -7.9293 & 0.06 & -0.84 & 0.05 & 0.35 & 1 & 11.196 & -0.747 & 0.95 & 0.49 & 0.07 & 1.57 & 9.91 & -0.04 & 0.77 & 5.8 & -70.74 \\
    7972-6101 & 316.7095 & 9.1974 & 0.04 & -1.43 & 0.01 & 0.16 & 1 & 11.269 & -1.116 & 0.24 & 0.22 & 0.04 & 1.9 & 9.87 & 0.1 & 0.71 & 5.26 & -74.11 \\
    7815-1901 & 317.5022 & 11.5106 & 0.02 & -1.33 & 0.02 & 0.36 & 1 & 9.964 & -1.952 & -0.08 & 0.09 & 0.08 & 1.69 & 9.71 & -0.06 & 0.58 & 1.01 & 51.18 \\
    8618-6104 & 319.8148 & 10.0706 & 0.07 & -0.96 & 0.01 & 0.18 & 1 & 11.206 & -1.21 & -0.4 & 0.53 & 0.04 & 1.67 & 9.93 & -0.02 & 0.97 & 5.51 & 48.98 \\
    8619-3704 & 324.2991 & 10.6638 & 0.08 & -1.49 & 0.01 & 0.11 & 1 & 11.181 & -1.387 & 1.06 & 0.32 & 0.01 & 1.85 & 9.83 & 0.03 & 0.33 & 4.86 & -79.31 \\
    7975-6103 & 324.7993 & 11.9393 & 0.08 & -2.89 & 0.01 & 0.13 & 1 & 11.179 & -1.065 & 0.76 & 0.3 & 0.01 & 1.83 & 9.87 & 0.06 & 0.72 & 5.11 & -16.02 \\
    8620-12703 & 330.8813 & 12.626 & 0.03 & -1.99 & 0.04 & 0.28 & 1 & 10.016 & -2.172 & 0.31 & 0.94 & 0.01 & 1.5 & 9.51 & -0.43 & 0.46 & 6.75 & 26.0 \\
\enddata
\tablecomments{ Table columns and other details are the same as Table \ref{tab: SFE_Table}. Typical errors (median error across all galaxies) : $\Delta$ TTYPE $= 0.63$, $\Delta P\_LTG = 0.01$,  $\Delta P\_S0 = 0.17$, $\Delta\log(\frac{M_\ast}{M_\odot}) = 0.02$, $\Delta\log(SFR) = 0.58$, $\Delta \lambda_{Re} = 0.03$, $\Delta H\alpha) = 0.01$, $\Delta D4000 = 0.01$, $\Delta \log(age) = 0.02$, $\Delta \log(z/z_{\odot}) = 0.02$. The full Table is available online.}
\end{splitdeluxetable*}

\begin{splitdeluxetable*}{ccccccccccBccccccccc}
\tabletypesize{\scriptsize}
\tablewidth{0pt} 
\tablecaption{Table of the control sample of SF-Sps. \label{tab:SFSp_Table}}
\tablehead{
\colhead{PLATEIFU} & \colhead{RA (J2000)}& \colhead{DEC (J2000)} & \colhead{z} & \colhead{TTYPE} & \colhead{P\_LTG} & \colhead{P\_S0} & \colhead{VC} & \colhead{$\log(\frac{M_\ast}{M_\odot})$} &
\colhead{$\log$(SFR)} \\
\colhead{} & \colhead{(degree)} &  \colhead{(degree)} & \colhead{} & \colhead{} & \colhead{} & \colhead{} & \colhead{} & \colhead{} & \colhead{($M_\odot$ yr$^{-1}$)} &
\colhead{$\log(\Sigma)$} & \colhead{$\lambda_{Re}$} & \colhead{H$\alpha$} & \colhead{D4000} & \colhead{log(age)} & \colhead{$\log(z/z_{\odot})$} & \colhead{B/T} & \colhead{Re} & \colhead{PA} \\
 \colhead{} & \colhead{} & \colhead{} & \colhead{} & \colhead{} & \colhead{} & \colhead{} & \colhead{} & \colhead{} & \colhead{} & \colhead{(Mpc$^{-2}$)} & \colhead{} &\colhead{$(10^{-16}erg/s/cm^2)$} & \colhead{} & \colhead{(yr)} & \colhead{} & \colhead{} & \colhead{(kpc)} & \colhead{(degree)}
} 
\colnumbers
\startdata 
12514-1902 & 200.4003 & 0.573 & 0.02 & 4.63 & 0.99 & 0.9 & 3 & 9.127 & -0.771 & -0.7 & 0.88 & 0.5 & 1.18 & 9.55 & -0.43 & 0.29 & 1.4 & 75.15 \\
    10843-12704 & 149.7077 & 0.8367 & 0.01 & 6.0 & 1.0 & 0.65 & 3 & 9.077 & -1.347 & -0.86 & 0.77 & 0.13 & 1.27 & 9.81 & -0.37 & 0.01 & 3.15 & -7.19 \\
    10843-12703 & 149.6573 & 0.8708 & 0.04 & 5.61 & 0.99 & 0.96 & 3 & 9.684 & -0.064 & -1.26 & 0.78 & 0.66 & 1.25 & 9.49 & -0.74 & 0.2 & 3.9 & 71.66 \\
    12071-3702 & 347.262 & 0.267 & 0.03 & 4.75 & 0.95 & 0.9 & 3 & 10.169 & 0.571 & -0.99 & 0.33 & 1.47 & 1.29 & 9.66 & -0.57 & 0.19 & 3.31 & -48.22 \\
    12071-9102 & 347.3455 & 1.0006 & 0.02 & 3.9 & 0.78 & 0.76 & 3 & 8.915 & -0.703 & -1.0 & 0.8 & 0.86 & 1.18 & 9.4 & -0.87 & 0.16 & 2.29 & -74.93 \\
    8654-12701 & 353.4445 & -0.6662 & 0.06 & 2.05 & 0.96 & 0.96 & 3 & 11.105 & 0.529 & 0.15 & 0.74 & 0.43 & 1.41 & 9.44 & -0.19 & 0.14 & 5.68 & 28.11 \\
    12696-9101 & 14.0174 & 0.261 & 0.05 & 2.82 & 0.83 & 0.98 & 3 & 10.217 & -0.129 & 0.56 & 0.96 & 0.25 & 1.22 & 9.66 & -0.21 & 0.12 & 3.98 & 51.03 \\
    12772-6104 & 17.5589 & 0.3659 & 0.05 & 2.59 & 0.95 & 0.92 & 3 & 10.418 & 0.195 & 0.66 & 0.86 & 0.33 & 1.48 & 9.71 & -0.23 & 0.17 & 3.32 & 42.53 \\
    12510-6102 & 202.5795 & -0.8707 & 0.04 & 2.26 & 0.86 & 0.93 & 3 & 9.698 & -0.233 & -0.88 & 0.61 & 0.77 & 1.3 & 9.6 & -0.81 & 0.62 & 2.39 & 10.35 \\
    12510-12701 & 202.4524 & -0.2942 & 0.02 & 6.37 & 1.0 & 0.99 & 3 & 9.816 & -0.467 & -0.51 & 0.9 & 0.16 & 1.27 & 9.36 & -0.65 & 0.74 & 6.02 & 31.86 \\
\enddata
\tablecomments{ Table columns and other details are the same as Table \ref{tab: SFE_Table}. Typical errors (median error across all galaxies): $\Delta$ TTYPE $= 0.68$, $\Delta P\_LTG = 0.02$,  $\Delta P\_S0 = 0.14$, $\Delta\log(\frac{M_\ast}{M_\odot}) = 0.04$, $\Delta\log(SFR) = 0.08$, $\Delta \lambda_{Re} = 0.02$, $\Delta H\alpha) = 0.04$, $\Delta D4000 = 0.02$, $\Delta \log(age) = 0.03$, $\Delta \log(z/z_{\odot}) = 0.05$. The full Table is available online.}
\end{splitdeluxetable*}

\section{Galaxy Images} \label{sec: imgs}
\begin{figure}
    \centering
    \includegraphics[width = 1\textwidth]{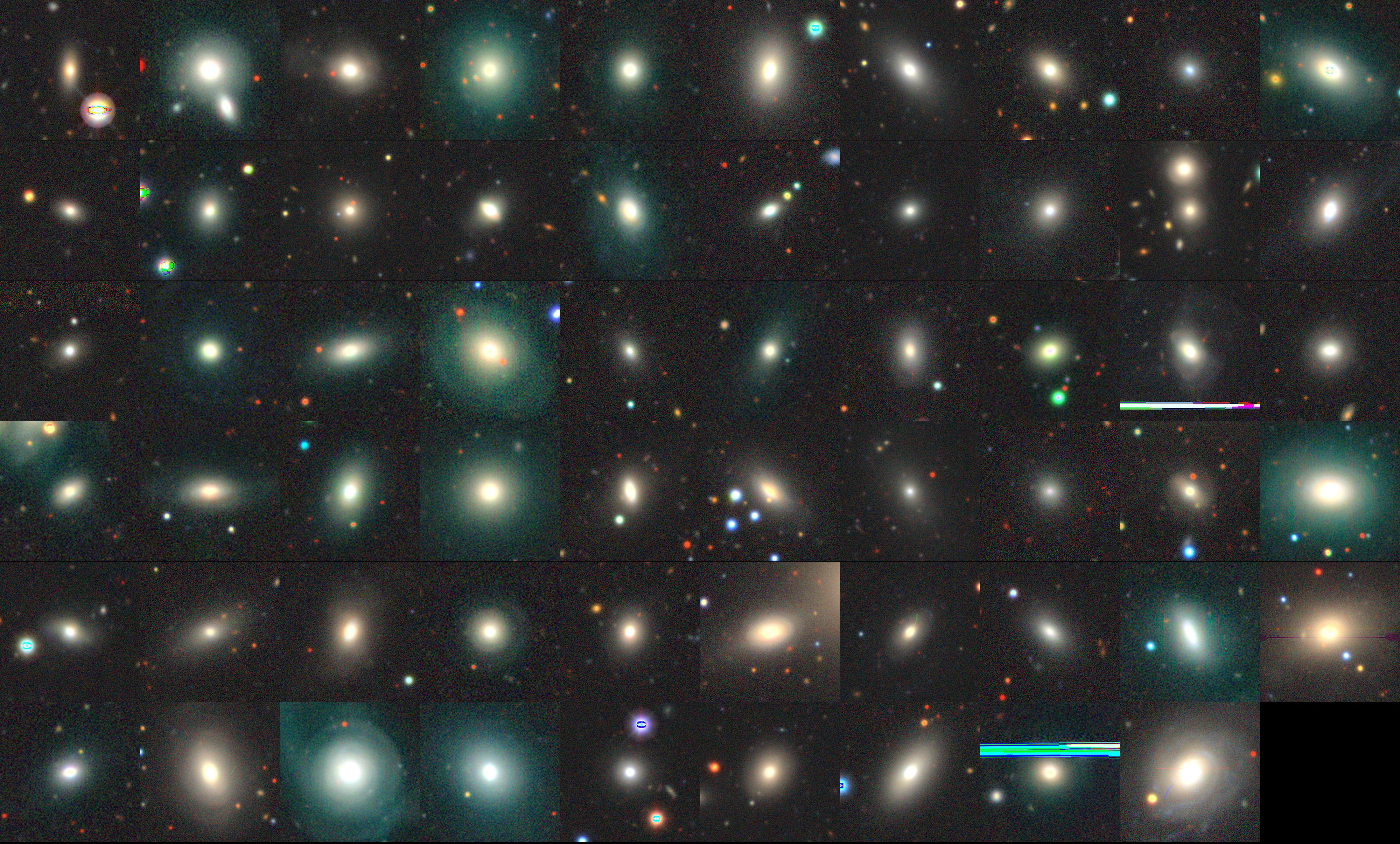}
    \caption{The DESI legacy imaging survey DR9 images of the primary sample of SF-Es. Each cutout is $60^{\prime\prime} \times 60 ^{\prime\prime}$. These images of SF-Es are in the same order (left to right row-wise) as presented in table~\ref{tab: SFE_Table}.}
    \label{fig:fig_A4}
\end{figure}

\begin{figure}
    \centering
    \includegraphics[width = 0.42\textwidth]{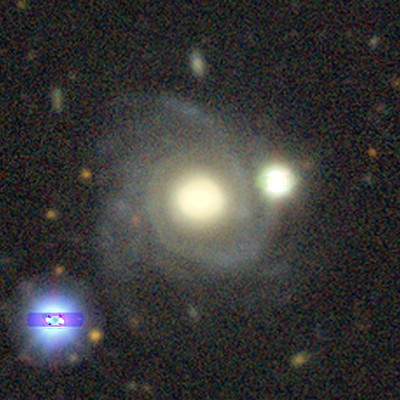}
    \includegraphics[width = 0.42\textwidth]{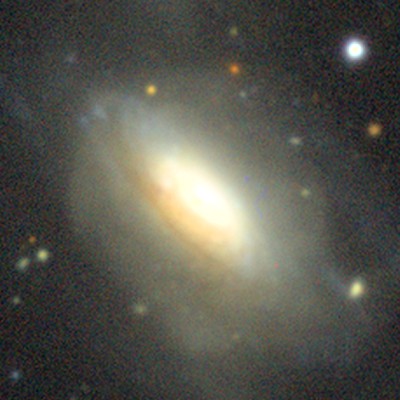}
    \caption{DESI legacy imaging survey DR10 images of two galaxies morphologically misclassified in the DL17 classification. These galaxies are probably spirals but are classified as ellipticals in DL17 likely due to the shallower imaging in SDSS, which was used to perform the morphological classification. This highlights the importance of deep imaging for accurate morphological classification of galaxies.  PLATEIFUs (left): 12068-12705 and (right): 8078-9102. Each image is $60^{\prime\prime} \times 60 ^{\prime\prime}$.}
    \label{fig:fig_A1}
\end{figure}

\begin{figure}
    \centering
    \includegraphics[width = 1\textwidth]{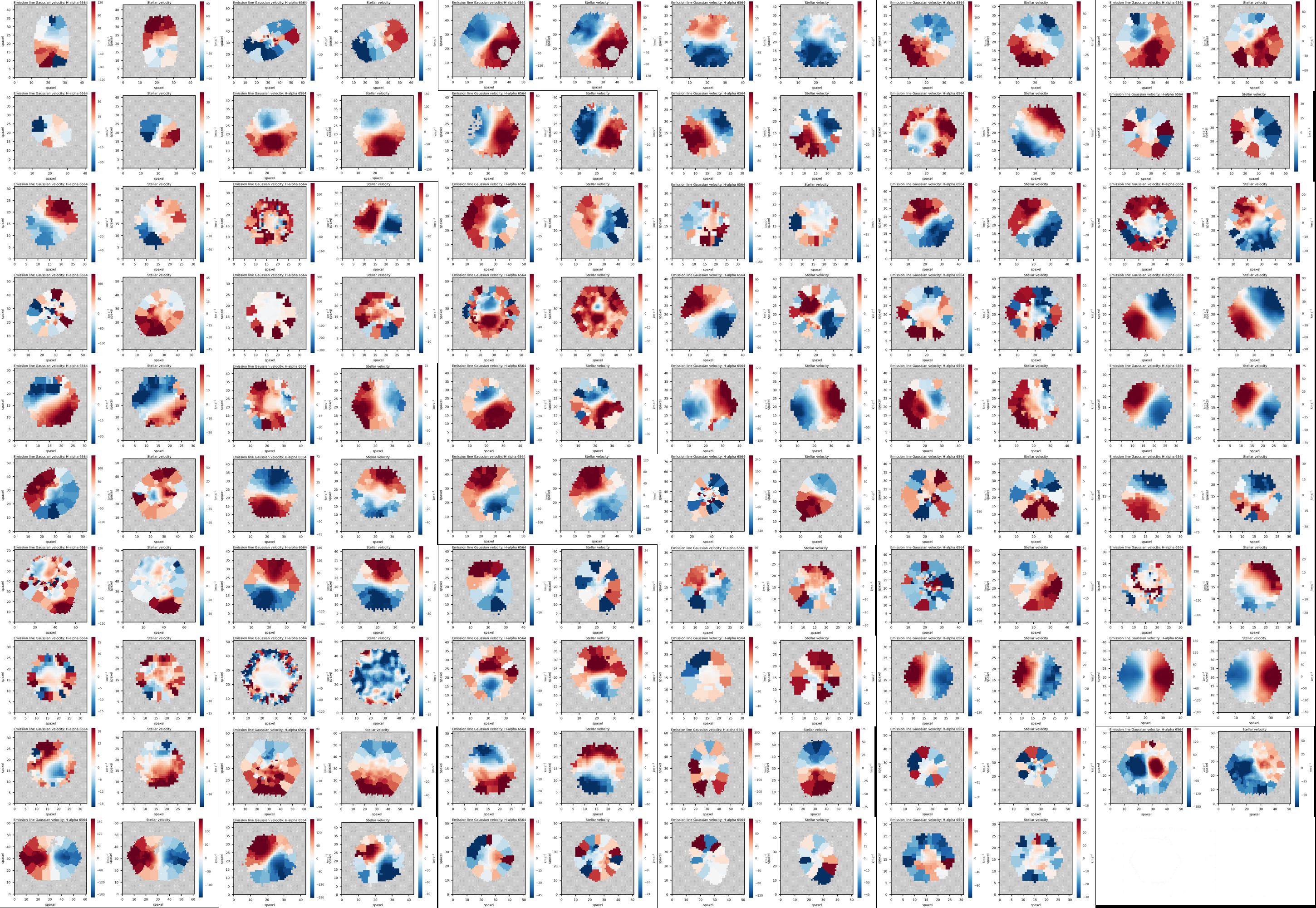}
    \caption{H$\alpha$ gas (left) and stellar (right) velocity maps of all galaxies in the primary sample of SF-Es. These maps of SF-Es are in the same order (left to right row-wise) as presented in table~\ref{tab: SFE_Table}.}
    \label{fig:fig_A2}
\end{figure}

\begin{figure}
    \centering
    \includegraphics[width = 1\textwidth]{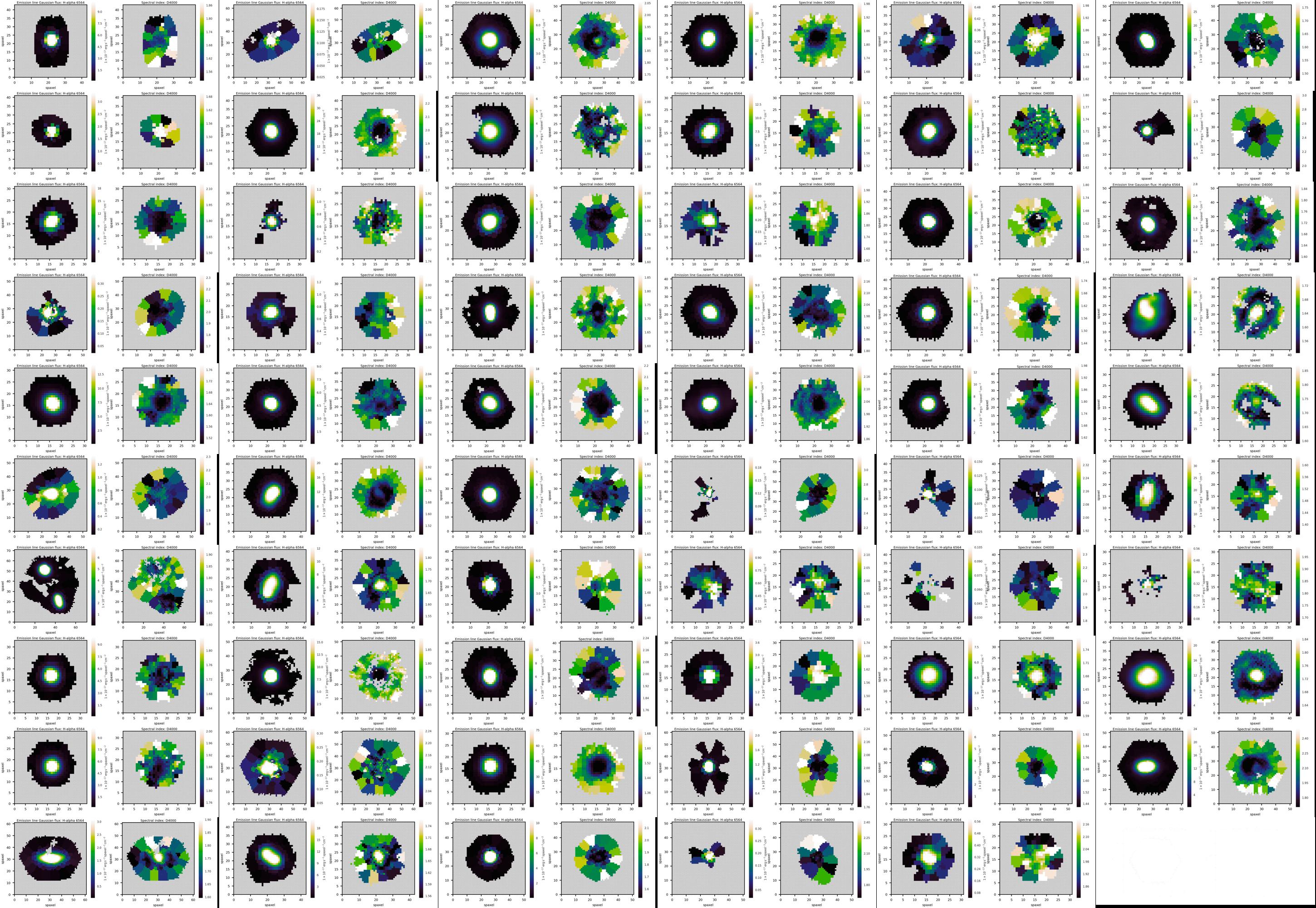}
    \caption{H$\alpha$ emission line Gaussian flux (left) and spectral index D4000 (right) maps of all galaxies in the primary sample of SF-Es. These maps of SF-Es are in the same order (left to right row-wise) as  presented in Table~\ref{tab: SFE_Table}.}
    \label{fig:fig_A3}
\end{figure}

Figure \ref{fig:fig_A4} shows image cutouts of all the galaxies in the primary sample of SF-Es. The two misclassified galaxies are displayed in Figure \ref{fig:fig_A1}. H$\alpha$ gas- and stellar-velocity maps are shown in Figure \ref{fig:fig_A2}, and Figure \ref{fig:fig_A3} represents the H$\alpha$ emission line flux and D4000 spectral index maps of SF-Es.

\bibliography{starFormingEs}{}
\bibliographystyle{aasjournal}



\end{document}